\documentclass[aps,twocolumn,pra,superscriptaddress,amsmath,showpacs,tightenlines]{revtex4}
\usepackage{epsfig,graphicx,times}
\usepackage{amstext}
\usepackage{amsmath}            
\usepackage{amssymb}            
\usepackage{graphicx}           
\usepackage{latexsym}
\usepackage{bm}
\begin{document}
\title{Entangled state engineering of vibrational modes in a multi-membrane optomechanical system}

\author{Xun-Wei Xu}
\affiliation{Institute of Microelectronics, Tsinghua University,
Beijing 100084, China}

\author{Yan-Jun Zhao}
\affiliation{Institute of Microelectronics, Tsinghua University,
Beijing 100084, China}

\author{Yu-xi Liu}
\email[]{yuxiliu@tsinghua.edu.cn}
\affiliation{Institute of Microelectronics, Tsinghua University,
Beijing 100084, China} \affiliation{Tsinghua National Laboratory for
Information Science and Technology (TNList), Tsinghua University,
Beijing 100084, China}

\date{\today}

\begin{abstract}
We propose a method to generate entangled states of the
vibrational modes of $N$ membranes which are coupled to a cavity mode via the
radiation pressure. Using sideband excitations, we show
that arbitrary entangled states of vibrational modes of different
membranes can be produced in principle by sequentially applying a
series of classical pulses with desired frequencies, phases and
durations. As examples, we show how to synthesize several
typical entangled states, for example, Bell states, NOON states, GHZ
states and W states. The environmental effect, information leakage,
and experimental feasibility are briefly discussed. Our proposal
can also be applied to other experimental setups of optomechanical systems, in
which many mechanical resonators are coupled to a common sing-mode cavity
field via the radiation pressure.
\end{abstract}

\maketitle

\section{Introduction}

Entanglement is  one of the most important resources for quantum
information processing~\cite{Nielsen}. Entanglement of the
internal freedoms of microscopic systems has already been prepared
experimentally~\cite{Raimond}, for example, the polarizations of
photons~\cite{Kwiat}, the electronic states of atoms~\cite{Hagley}
and spin states of ions~\cite{Sackett}. The entangled states in
macroscopic superconducting quantum systems have been experimentally
demonstrated~\cite{Martinis,Schoelkopf}. Recently, the quantum
properties of the mechanically vibrational modes are extensively
studied from external degrees of freedom of microscopic particles
(e.g., trapped ions ~\cite{Wineland2011}) to macroscopic
objects~\cite{physrep}.

The entanglement generation and arbitrary quantum state preparation
of vibrational modes in microscopic systems (e.g.,trapped ions) have
been studied in both experiments and theories (e.g., see
Refs.~\cite{Leibfried2003,PhysRep,SBZheng,LFWei}). Thus we would
like to know whether the entanglement can be generated in the
systems of macroscopic mechanical resonators. The research on the
coupling between superconducting quantum devices and macroscopic
mechanical resonators shows that arbitrary phonon states can be
produced in principle by using a proposed method~\cite{liuEuroph}
when a macroscopic mechanical resonator is coupled to a
superconducting qubits~\cite{OConnell}, also the generation of the
squeezed and entangled states of two vibrational modes has been
proposed by coupling two macroscopic mechanical resonators to the
superconducting quantum devices~\cite{feixue}. However, their
experimental realizations are still very challenge. The main
obstacle is whether the macroscopic mechanical resonator can be in
its ground state.

The ground state cooling~\cite{Gigan, Arcizet, Schliesser,
Wilson-Rae, MarquardtPRL, TeufelPRL, Thompson, Schliesser2008,
Rocheleau, Groblacher2009, Park, Schliesser2009, Teufel475,
Chan2009, Verhagen,jqliao} of the macroscopic mechanical resonators has
been theoretically studied and experimentally demonstrated by
coupling them to a cavity field via the radiation pressure. Thus the
optomichanical systems (see, reviews~\cite{Kippenberg, Marquardt})
provide a very good platform to study quantum mechanics at the
macroscopic scale. The stationary entanglement between mechanical
and cavity modes in the optomechanical systems has been
studied~\cite{Genes,Vitali07PRL,Vitali07PRA,AMari,HXMiao12}, and
such continuous variable entanglement can be used to implement
quantum teleportation~\cite{LTian06,SGHofer,SBarzanjeh12}.
Also both tripartite and bipartite entanglement between mechanical
modes and other degrees of freedom can be generated in the
optomechanical systems~\cite{yingdan,tianlin,MPaternostro07,CGenesNJP08,CGenes08,AXuereb12,MAbdi12}
or the hybrid optomechanical systems with an atomic
ensemble~\cite{Genes2008,Ian2008,DeChiara2011,Genes2011,LZhou11,LHSun,BRogers,KHammerer09,huijing,yuechang}
or a single-atom~\cite{MWallquist10,Barzanjeh11} inside the cavity.
Moreover, the cavity field mediated entanglement between two
macroscopic mechanical
resonators in the steady state ~\cite{Mancini2002,SPirandola06,Hartmann2008,Ludwig10,MBhattacharya08,MSchmidtNJP12,GVacantiNJP08,KBorkje11,LMazzola11,CJoshi12,MAbdiPRL12}
has also been theoretically explored. However, the coherent engineering of arbitrarily entangled phonon states of the macroscopic mechanical resonators is still an open question.

We have studied a deterministic method, which is different from the
proposals on measurement-based on phonon state
generation~\cite{pepper,kim}, to synthesize arbitrary nonclassical
phonon states in optomechanical systems~\cite{Xu_Liu}. Recent
studies show that many macroscopic mechanical resonators can be
coupled to a common single-mode cavity field~\cite{M1,M2,Tomadin,
Xuereb,Schmidt,holmes,Seok,Massel} via the radiation pressure, which
has been theoretically studied for selected entanglement
generation~\cite{Schmidt}, synchronization~\cite{holmes} and
mechanical analogue of nonlinear quantum optics~\cite{Seok} of many
mechanical modes, and also experimentally demonstrated tripartite
mixing for one cavity mode and two mechanical mode in the system
that a microwave cavity is coupled to two and more mechanical
resonators~\cite{Massel}. In such system, the single-mode cavity
field is a very good candidate to act as a data bus for information
transfer from one mechanical mode to
another one. Motivated by those researches,  we use
a multiple membrane optomechanical system~\cite{M1,M2,Tomadin, Xuereb}  as an example to study
entangled phonon state engineering, in optomechanical systems with the many mechanical resonators
coupled to a single-mode cavity field, by using the sideband excitations and and the single-photon effect~\cite{jqliao1} induced by the photon blockade. In particular, we will study detailed steps on engineering several typical entangled phonon states.

The paper is organized as follows. In Sec.~II, the theoretical
model of the optomechanical system with multiple membranes inside
a single-mode cavity is introduced. In Sec.~III, we
study a method to generate entangled states for the
system parameters within so-called Lamb-Dicke approximation. As examples, we show
how to generate Bell, NOON, GHZ and W states. In Sec.~IV, we study
entangled state engineering beyond the Lamb-Dicke
approximation for the strong single-photon optomechanical coupling.
Finally, discussions on experimental feasibility
and conclusions are given in Sec.~V.

\section{Theoretical model}
\subsection{Mode equations and transfer matrix theory}
The mode equations of optomechanical systems with one and two
membranes inside cavity have been studied by using the boundary conditions and the Helmholtz equations~\cite{M1,M2,Hartmann2008}.
However, the mode equations become harder and harder to be solved
with the increase of the membrane number. The transfer-matrix
method has been used widely in optics to analyze the propagation of
electromagnetic fields especially in multi-layer
structures~\cite{Born}. It has been used to
study the scattering problems in optomechanical
systems~\cite{Xuereb1,Xuereb2,Horsley}. For the completeness of the
paper, we will first derive intrinsic mode equations of the
optomechanical systems with $N$ membranes inside a cavity using the
transfer-matrix method.

\begin{figure}[tbp]
\includegraphics[bb=15 250 585 660, width=8 cm, clip]{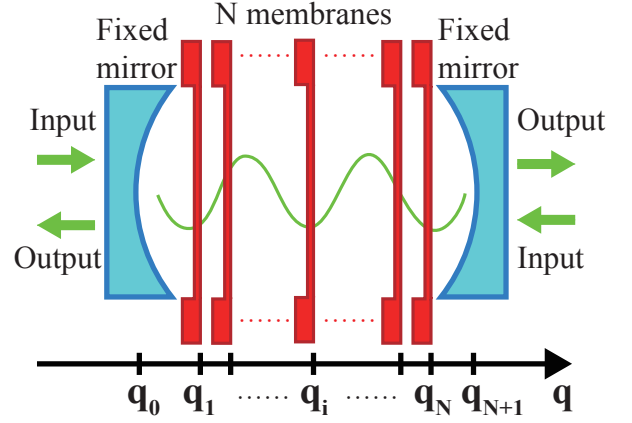}
\caption{Schematic diagram of an optomechanical system with $N$
mechanical membranes inside the cavity which is driven
by a classical field. Here, the two mirrors of the cavity are fixed.
$q_{0}$ and $q_{N+1}$ denote the positions of the cavity mirrors,
$q_{1},\cdots,q_{i},\cdots,q_{N}$ denote the positions of the $N$ membranes.}
\label{fig1}
\end{figure}

As a schematic diagram in Fig.~\ref{fig1}, we focus on an optomechanical system with
a cavity containing $N$ non-absorptive
membranes, that each has reflection coefficient $R$, mass $M_{i}$, position
$q_{i}$, and vibration frequency $\omega_{i}$ ($i=1,\cdots, N$). We assume that the thickness of each
vibrational membrane is much smaller than the wave length of the
cavity mode, so the electric susceptibility in the cavity can be
approximatively described by a sum of the dielectric permittivity
with $\delta$ functions~\cite{Spencer1972,Fader1985}
\begin{equation} \label{eq:1}
\varepsilon \left( x\right) =\varepsilon _{0}\left( 1+\frac{\zeta }{k}%
\sum_{i=1}^{N}\delta \left( x-q_{i}\right) \right),
\end{equation}
where $\varepsilon _{0}$ is the dielectric permittivity of vacuum,
$\zeta =2\sqrt{R/(1-R)}$, and $k=\omega/c$ is the wave vector of the
electric field with the mode frequency $\omega$ and the speed $c$ of
light in vacuum.

It is well known that the transfer matrix describing the electric
field through the empty space of
length $l$ is given by~\cite{Born}
\begin{equation} \label{eq:2}
M\left( k,l\right) =\left(
\begin{array}{cc}
\cos kl & \frac{1}{k}\sin kl \\
-k\sin kl & \cos kl%
\end{array}%
\right).
\end{equation}%
Let us now study the transfer matrix of the whole system by exploring the properties of
the electric fields cross a membrane. The
continuity of $E\left( x\right) $\ at the position of the $i$th
membrane (e.g. $x=q_{i}$) is give as
\begin{equation} \label{eq:3}
E\left( q_{i}^{+}\right)= E\left( q_{i}^{-}\right),
\end{equation}%
where $E\left( q_{i}^{-}\right)$ and $E\left( q_{i}^{+}\right)$ are
the notations of the left- and right-hand limits of $E\left(
x\right)$ when $x$ approaches $q_{i}$. Using the Helmholtz
equations, $ E^{\prime \prime }\left( x\right) =-\omega ^{2}\mu
_{0}\varepsilon \left( x\right) E\left( x\right)$,
the derivative relations of $E(x)$ in the left- and right-hand of the $i$th membrane at the position
$q_{i}$ is given as
\begin{equation} \label{eq:4}
E^{\prime }\left( q_{i}^{+}\right)=E^{\prime
}\left( q_{i}^{-} \right) -k\zeta E\left( q_{i}\right),
\end{equation}
with $E^{\prime }\left( x\right) = dE\left( x\right) /dx$.
Eqs.~(\ref{eq:3}) and (\ref{eq:4}) can be written in a matrix form
as
\begin{equation} \label{eq:5}
\left(
\begin{array}{c}
E\left( q_{i}^{+}\right)  \\
E^{\prime }\left( q_{i}^{+}\right)
\end{array}%
\right)=Q\left( k,\zeta \right) \left(
\begin{array}{c}
E\left( q_{i}^{-}\right)  \\
E^{\prime }\left( q_{i}^{-}\right)
\end{array}%
\right),
\end{equation}%
where
\begin{equation} \label{eq:6}
Q\left( k,\zeta \right) =\left(
\begin{array}{cc}
1 & 0 \\
-k\zeta  & 1%
\end{array}%
\right)
\end{equation}%
is the transfer matrix of the electric field through the $i$th membrane.
Using Eq.~(\ref{eq:2}) and (\ref{eq:5}),  the relation of
the electric fields at the left and right mirrors can be given by
\begin{equation}\label{eq:8}
\left(
\begin{array}{c}
E\left( q_{N+1}\right)  \\
E^{\prime }\left( q_{N+1}\right)
\end{array}%
\right)=X_{N}\left(
\begin{array}{c}
E\left( q_{0}\right)  \\
E^{\prime }\left( q_{0}\right)
\end{array}%
\right),
\end{equation}%
with the transfer matrix
\begin{eqnarray}\label{eq:9}
X_{N} &=&\left(
\begin{array}{cc}
x_{11} & x_{12} \\
x_{21} & x_{22}%
\end{array}\right) \\
&=&\prod_{i=1}^{N}\left[ M\left(
k,q_{i+1}-q_{i}\right) Q\left( k,\zeta\right) \right] M\left(
k,q_{1}-q_{0}\right)  . \nonumber
\end{eqnarray}%
Eq.~(\ref{eq:8}) can be rewritten as
\begin{eqnarray}
E\left( q_{N+1}\right)  &=&x_{11}E\left( q_{0}\right)
+x_{12}E^{\prime
}\left( q_{0}\right),   \label{eq:10}\\
E^{\prime }\left( q_{N+1}\right)  &=&x_{21}E\left( q_{0}\right)
+x_{22}E^{\prime }\left( q_{0}\right). \label{eq:11}
\end{eqnarray}%
If we assume that the electric field satisfies the standing wave boundary
conditions
\begin{equation}\label{eq:12}
E\left( q_{0}\right) =E\left( q_{N+1}\right) =0,
\end{equation}%
then we obtain the intrinsic mode equation%
\begin{equation} \label{eq:13}
x_{12}\left( k,q_{0},q_{1},\cdots ,q_{N+1}\right) =0.
\end{equation}

We further study the intrinsic mode equation in
Eq.~(\ref{eq:13}) by several concrete examples. For
$N=1$, the wave vector $k$ obeys the intrinsic mode equation%
\begin{equation}
C_{1,0}+\zeta C_{1,1}=0,
\end{equation}%
with%
\begin{eqnarray*}
C_{1,0} &=&\sin k\left( q_{2}-q_{0}\right),  \\
C_{1,1} &=&\sin k\left( q_{1}-q_{2}\right) \sin k\left(
q_{1}-q_{0}\right).
\end{eqnarray*}%
For $N=2$, the intrinsic mode equation for $k$ is
\begin{equation}
C_{2,0}+\zeta C_{2,1}+\zeta ^{2}C_{2,2}=0,
\end{equation}%
with%
\begin{eqnarray*}
C_{2,0} &=&\sin k\left( q_{3}-q_{0}\right),  \\
C_{2,1} &=&\sin k\left( q_{1}-q_{0}\right) \sin k\left(
q_{1}-q_{3}\right),\\
&&+\sin k\left( q_{2}-q_{0}\right) \sin k\left( q_{2}-q_{3}\right),  \\
C_{2,2} &=&\sin k\left( q_{1}-q_{0}\right) \sin k\left(
q_{2}-q_{1}\right) \sin k\left( q_{3}-q_{2}\right).
\end{eqnarray*}%
If there are $N$ membranes inside the cavity, the wave vector $k$
obeys the following equation
\begin{equation}
\sum_{i=0}^{N}\zeta ^{i}C_{N,i}=0,
\end{equation}%
where $C_{N,i}$ are functions of the $q_{i}$, $i=0,1,2,\cdots,N+1$.

\begin{figure}
\includegraphics[bb=57 237 524 580, width=8 cm, clip]{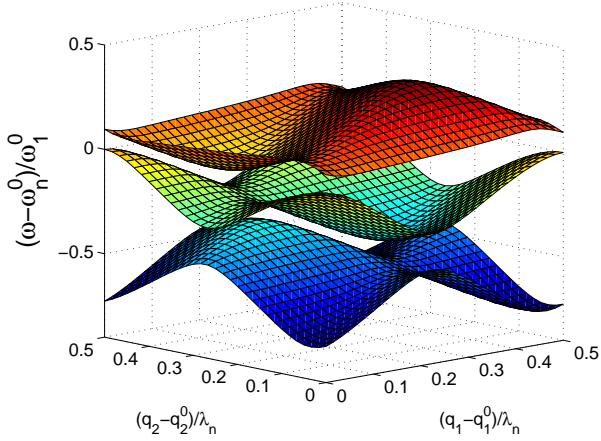}
\caption{The frequency shift of the intrinsic optical modes
$(\omega-\omega_{n}^{0})/\omega_{1}^{0}$ for two membranes in the cavity as
function of displacements of the membranes,
$(q_{1}-q_{1}^{0})/\lambda_{n}$ and $(q_{2}-q_{2}^{0})/\lambda_{n}$,
where $\lambda_{n}=2L/n$,  $\omega_{1}^{0}=\pi c/L$, $\omega_{n}^{0}=n\pi
c/L$, $q_{3}-q_{2}^{0}=q_{2}^{0}-q_{1}^{0}=q_{1}^{0}-q_{0}=L$. The
parameters are $R=0.7$, $n=10^5$.} \label{fig2}
\end{figure}

The frequency dependence of the cavity modes $\omega \left(
\left\{ q_{i}\right\} \right)$ on the positions of the $N$ membranes
can be obtained by solving the intrinsic mode equation in
Eq.~(\ref{eq:13}) numerically or analytically by using perturbation
theory. Here, $\{q_{i}\}$ is an abbreviation of $\{q_{1}, \cdots, q_{i}, \cdots,
q_{N}\}$ for the set of the positions of the membranes. For
example, in Fig.~\ref{fig2}, several intrinsic mode frequencies have
been numerically simulated and plotted as functions of the
displacements of the membranes when there are two vibrating
membranes inside the cavity. Analytically, the frequencies of the
cavity modes can be given as
\begin{eqnarray}
\omega \left( \left\{ q_{i}\right\} \right)  &=&\omega \left(
\left\{ q_{i}^{0}\right\} \right)  +\underset{i=1}{\overset{N}{\sum
}}g_{i}^{(1)}\left(
q_{i}-q_{i}^{0}\right)  \label{eq:18}  \\
&&+\underset{i,j=1}{\overset{N}{\sum }}g_{i,j}^{(2)}\left(
q_{i}-q_{i}^{0}\right) \left( q_{j}-q_{j}^{0}\right)+\cdots ,\notag
\end{eqnarray}
under the condition $(q_{i}-q_{i}^{0})/\lambda \ll 1$,  where
\begin{eqnarray}
g_{i}^{(1)}&=&\left[ \frac{\partial \omega \left( \left\{
q_{i}\right\} \right) }{\partial q_{i}}\right] _{\left\{
q_{i}=q_{i}^{0}\right\} }, \\
g_{i,j}^{(2)}&=&\frac{1}{2}\left[ \frac{\partial ^{2}\omega \left(
\left\{ q_{i}\right\} \right) }{\partial q_{i}\partial q_{j}}\right]
_{\left\{ q_{i}=q_{i}^{0},q_{j}=q_{j}^{0}\right\} },
\end{eqnarray}
$\lambda$ is the wave length of optical mode, and $q_{i}^{0}$
($i=1,\cdots, N$) is the position of the $i$th membrane when there is
no radiation pressure.

\subsection{Hamiltonian of the system}
Based on above discussions, the Hamiltonian of the optomechanical system with $N$ membranes inside the cavity can be written as~\cite{Law1995}
\begin{equation}
H =\hbar \omega \left( \left\{ q_{i}\right\} \right) a^{\dag }a+%
\underset{i=1}{\overset{N}{\sum }}\left[ \frac{p_{i}^{2}}{2M_{i}}+\frac{1}{2}%
M_{i}\omega _{i}^{2}\left( q_{i}-q_{i}^{0}\right) ^{2}\right], \label{eq:21}
\end{equation}
where $a$ ($a^{\dagger }$) is the annihilation (creation) operator
of the cavity field, $p_{i}$ is the momentum of the $i$th vibrational
membrane. Substituting Eq.~(\ref{eq:18}) into Eq.~(\ref{eq:21}), we have
\begin{eqnarray}
H &=&\hbar \omega \left( \left\{ q_{i}^{0}\right\} \right) a^{\dag }a+%
\underset{i=1}{\overset{N}{\sum }}\left[ \frac{p_{i}^{2}}{2M_{i}}+\frac{1}{2}%
M_{i}\omega _{i}^{2}\left( q_{i}-q_{i}^{0}\right) ^{2}\right]   \notag \\
&&+\hbar a^{\dag }a\underset{i=1}{\overset{N}{\sum
}}g_{i}^{(1)}\left(
q_{i}-q_{i}^{0}\right)   \\
&&+\hbar a^{\dag }a\underset{i,j=1}{\overset{N}{\sum
}}g_{i,j}^{(2)}\left( q_{i}-q_{i}^{0}\right) \left(
q_{j}-q_{j}^{0}\right)+\cdots   \notag .
\end{eqnarray}
The third and fourth term are the linear and quadratic interactions
between the cavity mode and the vibrational modes of the membranes.

Hereafter, we only consider that the frequency shift of the cavity mode
is linearly dependent on the membranes' displacements. We also
assume that the cavity is driven by an external field with the
frequency $\omega _{d}$ and the phase $\phi_{d}$.  Thus, we have the
Hamiltonian of the driven system as below
\begin{eqnarray}\label{eq:22}
H_{d} &=&\hbar \omega _{a}a^{\dag }a+\underset{i=1}{\overset{N}{\sum
}}\hbar \omega _{i}b_{i}^{\dag }b_{i} +\hbar a^{\dag }a \sum_{i=1}^{N} g_{i}\left(
b_{i}^{\dag }+b_{i}\right)   \nonumber \\
&&+\hbar \Omega \left[ a^{\dag }e^{-i\left( \omega _{d}t+\phi
_{d}\right) }+ae^{i\left( \omega _{d}t+\phi _{d}\right) }\right].
\end{eqnarray}%
$\Omega$ denotes the Rabi frequency of the driven field.
We assume that both the frequency $\omega _{d}$ and the phase $\phi_{d}$ are
controllable parameters such that they can be chosen
as different values in the steps of the state preparation described
below. For the simplicity, the frequency of the cavity mode is denoted as
$\omega_{a}\equiv \omega \left( \left\{ q_{i}^{0}\right\} \right)$, and the coupling strength between the cavity field and the $i$th membrane is simply written as $g_{i}$. The operators of
the membranes are rewritten by the annihilation and creation
operators $b_{i}=\sqrt{M_{i}\omega_{i}/2\hbar}(q_{i}-q_{i}^{0})+ip_{i}/\sqrt{2\hbar M_{i}\omega_{i}}$ and $b_{i}^{\dagger}=\sqrt{M_{i}\omega_{i}/2\hbar}(q_{i}-q_{i}^{0})-ip_{i}/\sqrt{2\hbar M_{i}\omega_{i}}$. We now applying a unitary transformation to Eq.~(\ref{eq:22})
\begin{equation}
U=\exp \left( a^{\dag }a\underset{i=1}{\overset{N}{\sum }}\left[ \frac{g_{i}%
}{\omega _{i}}\left( b_{i}^{\dag }-b_{i}\right) \right] \right),
\end{equation}%
then the Hamiltonian in Eq.~(\ref{eq:22}) becomes into
\begin{eqnarray}
H_{\rm eff} &=&\hbar (\omega_{a}-\Delta _{0}a^{\dag }a)a^{\dag
}a+\hbar
\underset{i=1}{\overset{N}{\sum }}\omega _{i}b_{i}^{\dag }b_{i} \label{eq:24}  \\
&+&\hbar \Omega a^{\dag }\exp\left[{\underset{i=1}{\overset{N}{\sum
}}\eta _{i}\left( b_{i}^{\dag }-b_{i}\right)-i\left( \omega
_{d}t+\phi _{d}\right) }\right]+\rm {H.c.},\notag
\end{eqnarray}%
where $\Delta _{0}={\sum }^{N}_{i=1} (g_{i}^{2}/\omega _{i})$
characterizes the nonlinearity of the cavity field induced by the
vibrational membranes, thus the nonlinearity increases with the
increase of the number of the membranes. We call $\eta
_{i}=g_{i}/\omega _{i}$ as the Lamb-Dicke parameter in analog
to the trapped ions~\cite{Leibfried2003}. Both the strong
optomechanical coupling and many mechanical resonator make the nonlinear term $\Delta_{0}$
guarantee the photon blockade, in this case, the driving field can be assumed to be
coupled to two lowest energy levels $\left\vert 0\right\rangle
$ and $\left\vert 1\right\rangle $ of the cavity field. Thus, the
cavity field is confined to the states $|0\rangle$ and $|1\rangle$,
and the Hamiltonian in Eq.~(\ref{eq:24}) becomes into
\begin{eqnarray}\label{eq:25}
H_{\rm two} &=&\hbar \frac{\omega_0}{2}\sigma _{z}+\hbar
\sum_{i=1}^{N}\omega_{i}b_{i}^{\dag }b_{i}   \\
&+&\hbar \Omega \sigma _{+}\exp\left[\sum_{i=1}^{N}\eta_{i}(
b_{i}^{\dag }-b_{i})-i\left( \omega _{d}t+\phi
_{d}\right)\right]+{\rm H.c.}, \notag
\end{eqnarray}%
by using the operators $\sigma _{z}=\left\vert 1\right\rangle
\left\langle 1\right\vert -\left\vert 0\right\rangle \left\langle
0\right\vert $ and $\sigma _{+}=\left\vert 1\right\rangle
\left\langle 0\right\vert $. Here, $\omega_0 =\omega _{a}-\Delta
_{0}$. Hereafter, we denote the photon number states as $\left\vert
1\right\rangle \equiv \left\vert e\right\rangle $, $\left\vert
0\right\rangle \equiv \left\vert g\right\rangle $. The Hamiltonian
in Eq.~(\ref{eq:25}) can be further written as
\begin{equation}
H_{\rm two}=H_{0}+H_{\rm int},
\end{equation}%
with
\begin{equation}
H_{0}=\hbar \frac{\omega_{0}}{2}\sigma _{z}+\sum_{i=1}^{N}\hbar
\omega _{i}b_{i}^{\dag }b_{i},
\end{equation}%
and
\begin{eqnarray}\label{eq:29}
H_{\rm int} &=&\hbar \Omega \sigma _{+}e^{-i\left( \omega _{d}t+\phi
_{d}\right) }\prod_{i=1}^{N}\left[ e^{-\frac{1}{2}\eta _{i}^{2}} \sum_{j_{i},k_{i}=0}^{+\infty
}\frac{\left( -1\right)
^{k_{i}}\eta _{i}^{j_{i}+k_{i}}}{j_{i}!k_{i}!}\right.   \notag \\
&&\left. \times b_{i}^{\dag j_{i}}b_{i}^{k_{i}}%
\right] +{\rm H.c.}.
\end{eqnarray}%
From Eq.~(\ref{eq:29}), we find that $|k_{i}-j_{i}|$ phonons can be
created ($ k_{i}>j_{i}$) or annihilated ($k_{i}<j_{i}$) from the
$i$th membrane when one photon is annihilated in the cavity with the
assistance of the external field. Below, we will show entangled state engineering for different vibrational modes of the
membranes for two cases with or without Lamb-Dicke approximation.

\section{Engineering entangled states with Lamb-Dicke approximation}

We first study the entanglement engineering under the Lamb-Dicke
approximation condition $g _{i}/\omega _{i} \ll 1$ as for the
trapped ion case~\cite{Leibfried2003}, thus the Hamiltonian in
Eq.~(\ref{eq:29}) can be written as
\begin{equation}
H_{\rm int}=\hbar \Omega \sigma _{+}e^{-i\left( \omega _{d}t+\phi
_{d}\right) } \left[ 1+\sum_{i=1}^{N}\eta _{i}\left( b_{i}^{\dag
}-b_{i}\right) \right] +{\rm H.c.},
\end{equation}
up to the first order of $\eta _{i}$.
In the interaction picture with $V=\exp(iH_{0}t/\hbar)H_{\rm
int}\exp(-iH_{0}t/\hbar)$, we have
\begin{eqnarray}
V&=&\hbar \Omega \sigma _{+}e^{-i\phi _{d}}\left[ e^{-i\Delta
_{c}t}+\sum_{i=1}^{N}\eta _{i}\left( b_{i}^{\dag }e^{-i\Delta
_{b}^{i}t}-b_{i}e^{-i\Delta _{r}^{i}t}\right) \right] \nonumber \\
&&+{\rm H.c.},
\end{eqnarray}%
with $\Delta _{c}=\omega _{d}-\omega_{0} $, $\Delta _{b}^{i}=\omega
_{d}-\omega_{0} -\omega _{i}$, and $\Delta _{r}^{i}=\omega
_{d}-\omega_{0} +\omega _{i}$. If the system satisfies the
resonant condition either $\Delta_{c}=0$ or $\Delta_{b}^{i}=0$ or
$\Delta_{r}^{i}=0$,  and also the driving field is not very strong,
then we have
\begin{equation}\label{eq:30}
V=\hbar \Omega \times \left\{ \begin{array}{ll}
\sigma _{+}e^{-i\phi _{c}}+{\rm H.c.} & \omega _{d}=\omega_{0}, \\
\eta _{i}\sigma _{+}b_{i}^{\dag }e^{-i\phi _{b}^{i}}+{\rm H.c.} & \omega
_{d}=\omega_{0} +\omega _{i}, \\
\eta _{i}\sigma _{+}b_{i}e^{-i\phi _{r}^{i}}+{\rm H.c.} & \omega
_{d}=\omega_{0}-\omega_{i}.%
\end{array} \right.
\end{equation}
with the rotating wave approximation. For convenience, the minus
sign before $\eta _{i}\sigma _{+}b_{i}e^{-i\phi _{r}^{i}}$\ is
absorbed by the phase $\phi _{r}^{i}$.

\subsection{The time evolution operators}

From the Schr\"{o}dinger equation, the wave function of the system
at the time $t$ can be written as
\begin{equation}
\left\vert \psi \left( t\right) \right\rangle =U\left( t\right)
\left\vert \psi \left( 0\right) \right\rangle,
\end{equation}%
where $U\left( t\right) =\exp \left( -iVt/\hbar \right) $ is
the time evolution operator~\cite{LFWei}. By using the completeness relation
\begin{equation}  \label{eq:34}
\sum_{\left\{ m_{j}\right\} =0}^{+\infty }\sum_{s=g}^{e}\left\vert
s,\left\{ m_{j}\right\} \right\rangle \left\langle s,\left\{
m_{j}\right\} \right\vert =I,
\end{equation}%
the time evolution operator can be written as
\begin{equation}
U\left( t\right) =\sum_{\left\{ m_{j}\right\} =0}^{+\infty
}\sum_{s=g}^{e}U\left( t\right) \left\vert s,\left\{ m_{j}\right\}
\right\rangle \left\langle s,\left\{ m_{j}\right\} \right\vert,
\end{equation}
where $|\{m_{i}\}\rangle$ is an abbreviation of the state $|m_{1}\rangle\otimes \cdots \otimes |m_{i}\rangle \otimes \cdots \otimes|m_{N}\rangle\equiv |m_{1},\cdots, m_{N}\rangle$ for $N$ membranes. Here $\left\vert
s,\left\{ m_{j}\right\} \right\rangle$ implies that the cavity field is in the state $s$ ($s=e$ or $s=g$)
and there are $m_{j}$ phonons in the $j$th membranes. $\{m_{j}\}$ denotes a number series, that is,  $\{ m_{j}\}\equiv m_{1},m_{2},\cdots,m_{N}$.

If the frequency of the driving field is resonant with the
red-sideband excitation corresponding to the frequency of the $i$th membrane, i.e.,
$\omega _{d}=\omega_0 -\omega _{i}$,
then the Hamiltonian in Eq.~(\ref{eq:30}) becomes
\begin{equation}
V_{m_{i}}^{i,r}=\hbar \Omega \eta _{i}\sigma _{+}b_{i}e^{-i\phi
_{r}^{i}}+{\rm H.c.}.
\end{equation}%
In this case, the time evolution operator is given as
\begin{equation}
U_{m_{i}}^{i,r}\left( t\right) =\sum_{ m_{i} =0}^{+\infty }
\widetilde{U}_{m_{i}}^{i,r}\left( t\right)\sum_{\left\{ m_{j}\right\} =0}^{+\infty } \left(\left\vert \left\{
m_{j}\right\} \right\rangle \left\langle \left\{ m_{j}\right\}
\right\vert \right) _{j\neq i},
\end{equation}%
where
\begin{eqnarray}
\widetilde{U}_{m_{i}}^{i,r}\left( t\right)  &=&\cos \left( \Omega
_{m_{i}}^{i} t\right) \left\vert g,m_{i}\right\rangle \left\langle
g,m_{i}\right\vert   \nonumber \\
&&-ie^{-i\phi _{r}^{i}}\sin \left( \Omega _{m_{i}}^{i} t\right)
\left\vert
e,m_{i}-1\right\rangle \left\langle g,m_{i}\right\vert   \nonumber \\
&&+\cos \left( \Omega _{m_{i}+1}^{i} t\right) \left\vert
e,m_{i}\right\rangle
\left\langle e,m_{i}\right\vert   \nonumber \\
&&-ie^{i\phi _{r}^{i}}\sin \left( \Omega _{m_{i}+1}^{i} t\right)
\left\vert g,m_{i}+1\right\rangle \left\langle e,m_{i}\right\vert
\end{eqnarray}%
with the Rabi frequencies
\begin{equation}
\Omega _{m_{i}}^{i}=\Omega \eta _{i}\sqrt{m_{i}}\,,\;\;\;\;\; \Omega
_{m_{i}+1}^{i}=\Omega \eta _{i}\sqrt{m_{i}+1}.
\end{equation}

When the frequency of the driving field is resonant with the
blue-sideband excitation corresponding to the frequency of the $i$th membrane, i.e.,
$\omega _{d}=\omega_0 +\omega _{i}$,
the Hamiltonian in Eq.~(\ref{eq:30}) becomes
\begin{equation}
V_{m_{i}}^{i,b}=\hbar \Omega \eta _{i}\sigma _{+}b_{i}^{\dag
}e^{-i\phi _{b}^{i}}+{\rm H.c.}.
\end{equation}%
The time evolution operator of the blue-sideband excitation is
\begin{equation}
U_{m_{i}}^{i,b}\left( t\right) =\sum_{ m_{i}=0}^{+\infty }
\widetilde{U}_{m_{i}}^{i,b}\left( t\right) \sum_{\left\{ m_{j}\right\} =0}^{+\infty } \left( \left\vert \left\{
m_{j}\right\} \right\rangle \left\langle \left\{ m_{j}\right\}
\right\vert \right) _{j\neq i}
\end{equation}%
with
\begin{eqnarray}
\widetilde{U}_{m_{i}}^{i,b}\left( t\right)  &=&\cos \left( \Omega
_{m_{i}+1}^{i} t\right) \left\vert g,m_{i}\right\rangle \left\langle
g,m_{i}\right\vert   \nonumber \\
&&-ie^{-i\phi _{b}^{i}}\sin \left( \Omega _{m_{i}+1}^{i} t\right)
\left\vert
e,m_{i}+1\right\rangle \left\langle g,m_{i}\right\vert   \nonumber \\
&&+\cos \left( \Omega _{m_{i}}^{i} t\right) \left\vert
e,m_{i}\right\rangle \left\langle e,m_{i}\right\vert \left\langle
e,m_{i}\right\vert   \nonumber
\\
&&-ie^{i\phi _{b}^{i}}\sin \left( \Omega _{m_{i}}^{i} t\right) \left\vert
g,m_{i}-1\right\rangle \left\langle e,m_{i}\right\vert.
\end{eqnarray}

Finally, if the cavity is driven by the classical field with the
frequency
$\omega _{d}=\omega_0$,
then the carrier process is switched on, and the Hamiltonian in Eq.~(\ref{eq:30}) is given as
\begin{equation}
V^{c}=\hbar \Omega \sigma _{+}e^{-i\phi _{c}}+ {\rm H.c.}.
\end{equation}
The time evolution operator of the carrier process is given as
\begin{equation}
U^{c}\left( t\right) =\widetilde{U}
^{c}\left( t\right)\sum_{\left\{ m_{j}\right\} =0}^{+\infty } \left\vert \left\{ m_{j}\right\}
\right\rangle \left\langle \left\{ m_{j}\right\} \right\vert
\end{equation}
and
\begin{eqnarray}
\widetilde{U}^{c}\left( t\right)  &=&\left[ \cos \left( \Omega
t\right) \left\vert g\right\rangle -ie^{-i\phi _{c}}\sin \left(
\Omega t\right)
\left\vert e\right\rangle \right] \left\langle g\right\vert   \nonumber \\
&&+\left[ \cos \left( \Omega t\right) \left\vert e\right\rangle
-ie^{i\phi _{c}}\sin \left( \Omega t\right) \left\vert
g\right\rangle \right] \left\langle e\right\vert.
\end{eqnarray}
Based on above three different processes, we can in principle engineer any
kind of entangled phonon states. However, below we will only show the
engineering of several typical entangled phonon states by controlling
the evolution time and the frequency of the classical driving field.

\subsection{Generation of Bell and NOON states of two mechanical modes}

In this section, we study the generation of entangled
phonon states
\begin{equation}
\left\vert \varphi \right\rangle =\frac{1}{\sqrt{2}}\left( \left\vert
N,0\right\rangle +\left\vert 0,N\right\rangle \right) ,
\end{equation}%
of two vibrational modes when there are two membranes inside the cavity. Here
$\left\vert \varphi \right\rangle$  denotes the Bell state when $N=1$. However $\left\vert \varphi \right\rangle$ with $N\geq 2$ represents the NOON state
which plays an important role in quantum metrology~\cite{Bollinger}. $|N,0\rangle$ means
$N$ phonons in mode one and zero phonon in mode two. Our state generation below starts
from the initial state
$\left\vert \psi \left( t_{0} \right) \right\rangle =\left\vert
g,0,0\right\rangle$ of the whole system.

We now show how to generate a Bell state. First, a driving field
satisfying the carrier process is applied to the cavity,
then after the interaction time $\Delta t_{1}=\pi /2\Omega $,
the system state at the time $t_{1}=t_{0}+\Delta t_{1}$ is
\begin{equation}
\left\vert \psi \left( t_{1}\right) \right\rangle =\left\vert
e,0,0\right\rangle,
\end{equation}
here a global phase has been neglected.  Second, the frequency of
the driving field is turned to the red sideband corresponding to the
frequency of the first membrane, i.e., $\omega _{d}=\omega_0 -\omega _{1}$. With an
evolution time $\Delta t_{2}$, the system evolves to
\begin{equation}
\left\vert \psi \left( t_{2}\right) \right\rangle =\left( 1-\left\vert
C_{1,0}\right\vert ^{2}\right) ^{1/2}\left\vert e,0,0\right\rangle
+C_{1,0}\left\vert g,1,0\right\rangle
\end{equation}%
at the time $t_{2}=t_{1}+\Delta t_{2}$ with the parameter
\begin{equation}
C_{1,0}=-ie^{i\phi _{r}^{1}}\sin \left( \Omega _{1}^{1}\Delta t_{2}\right).
\end{equation}%
If the time duration and the phase of the driving field are chosen
as $\Delta t_{2}=\pi /4\Omega _{1}^{1}$, phase $\phi _{r}^{1}=\pi
/2$, then $C_{1,0}=1/\sqrt{2}$, and the state of the system becomes
into
\begin{equation}
\left\vert \psi \left( t_{2}\right) \right\rangle =\frac{1}{\sqrt{2}}%
\left\vert e,0,0\right\rangle +\frac{1}{\sqrt{2}}\left\vert
g,1,0\right\rangle.
\end{equation}
In the third step,  the frequency of the driving
field is tuned to the red sideband corresponding to the frequency of the second membrane, i.e., $\omega
_{d}=\omega_0 -\omega _{2}$. With the evolution time $\Delta t_{3}$,
the system evolves into
\begin{eqnarray}
\left\vert \psi \left( t_{3}\right) \right\rangle &=&\frac{1}{\sqrt{2}}\left[
\left( 1-\left\vert C_{0,1}\right\vert ^{2}\right) ^{1/2}\left\vert
e,0,0\right\rangle +C_{0,1}\left\vert g,0,1\right\rangle \right]  \nonumber \\
&&+\frac{1}{\sqrt{2}}\left\vert g,1,0\right\rangle
\end{eqnarray}%
at the time $t_{3}=t_{2}+\Delta t_{3}$ with the parameter
\begin{equation}
C_{0,1}=-ie^{i\phi _{r}^{2}}\sin \left( \Omega _{1}^{2}\Delta t_{3}\right).
\end{equation}%
If we choose the time duration $\Delta t_{3}=\pi /2\Omega _{1}^{2}$
and phase $\phi
_{r}^{2}=\pi /2$, then the state of the system becomes%
\begin{equation}
\left\vert \psi \left( t_{3}\right) \right\rangle =|g\rangle \otimes \frac{1}{\sqrt{2}} \left(
\left\vert 0,1\right\rangle +\left\vert 1,0\right\rangle \right).
\end{equation}%
Thus the system is deterministically prepared to a product state of the Bell state of two mechanical resonators and the ground state $\left\vert g\right\rangle $ of the cavity field.

\begin{figure}
\includegraphics[bb=105 520 500 640, width=9 cm, clip]{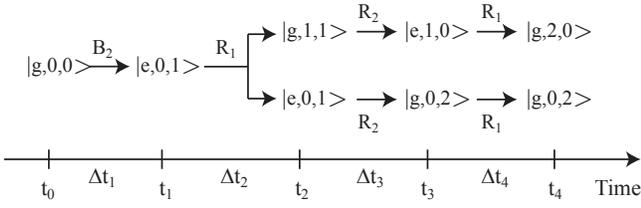}
\caption{Schematic diagrams for preparing NOON states $\left(
\left\vert 2,0\right\rangle +\left\vert 0,2\right\rangle
\right)/\sqrt{2}$. The letters $R$, $B$ and $C$ represent the
process that the driving field is tuned to the
red-sideband, blue-sideband or carrier processes, respectively. Each
number of the subscript denotes the sideband excitations corresponding to the
frequency of the $i$th membrane.} \label{fig3}
\end{figure}

We now describe a detailed steps for preparing an arbitrary NOON state of two vibrational
membranes by the simplest example with $N=2$.
First, a driving field is applied to the cavity
with the blue sideband excitation corresponding to frequency of the second
membrane, i.e. $\omega _{d}=\omega_0 +\omega _{2}$, with an
evolution time $\Delta t_{1}=\pi /2\Omega _{1}^{2}$ and choosing the
phase $\phi _{b}^{2}=3\pi /2$, the system evolves to
\begin{equation}
\left\vert\psi \left( t_{1}\right) \right\rangle =\left\vert
e,0,1\right\rangle
\end{equation}
at the time $t_{1}=t_{0}+\Delta t_{1}$.  Second, the
frequency of the driving field is tuned to the red sideband
excitation corresponding to the frequency of the first membrane, i.e.,
$\omega _{d}=\omega_0 -\omega_{1}$. With an evolution time $ \Delta t_{2}=\pi /4\Omega
_{1}^{1}$ and choosing the phase $\phi _{r}^{1}=\pi /2$,
the state of the system becomes%
\begin{equation}
\left\vert \psi \left( t_{2}\right) \right\rangle
=\frac{1}{\sqrt{2}}(\left\vert g,1,1\right\rangle+ \left\vert
e,0,1\right\rangle).
\end{equation}
at the time $t_{2}=t_{1}+\Delta t_{2}$. Third, the
frequency of the driving field is tuned to the red sideband
excitation corresponding to the frequency of the second membrane, i.e., $\omega _{d}=\omega_0 -\omega
_{2}$, with the time duration $\Delta t_{3}$ satisfying
\begin{equation} \label{eq:62}
\sin(\Omega^{2}_{1} \Delta  t_{3})=\pm 1, \sin(\Omega^{2}_{2} \Delta
t_{3})=\pm 1,
\end{equation}
in the infinite approximation for $\Omega^{2}_{2}/\Omega^{2}_{1}=\sqrt{2}$,
which is an irrational number (see the Appendix B), the state of
the whole system becomes
\begin{equation}
\left\vert \psi \left( t_{3}\right) \right\rangle
=\frac{1}{\sqrt{2}}( \left\vert e,1,0\right\rangle +\left\vert
g,0,2\right\rangle),
\end{equation}
at the time $t_{3}=t_{2}+\Delta t_{3}$. In the last step,  the
driving field is tuned to the red sideband excitation corresponding to the frequency of the first
membrane, i.e., $ \omega _{d}=\omega_0 -\omega _{1}$. With an evolution
time $ \Delta t_{4}=\pi /2\Omega^{1}_{2}$ and the phase $\phi
_{r}^{1}=\pi /2$, the state of the system becomes
\begin{equation}
\left\vert \psi \left( t_{4}\right) \right\rangle
=|g\rangle\otimes \frac{1}{\sqrt{2}}(\left\vert 2,0\right\rangle +\left\vert
0,2\right\rangle),
\end{equation}
at the time $t_{4}=t_{3}+\Delta t_{4}$. Thus the NOON state for
$N=2$ is generated with the cavity field in its ground state
$\left\vert g\right\rangle$. We have given a schematic diagram to
summary all steps for generating the NOON state with $N=2$ in
Fig.~\ref{fig3}.


\begin{figure}
\includegraphics[bb=15 350 555 635, width=9 cm, clip]{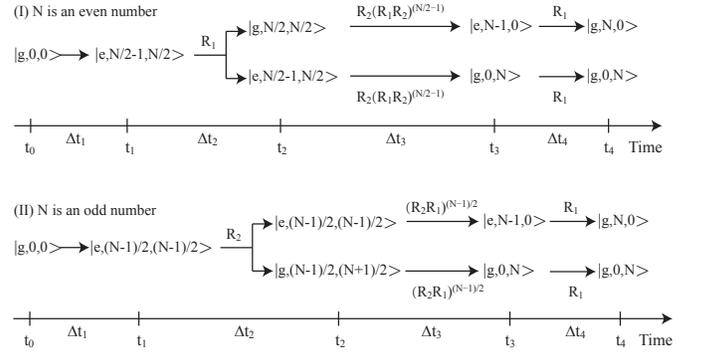}
\caption{Schematic diagram for preparing NOON states when (I) $N$ is
an even number and (II) $N$ is an odd number. The
numbers in the superscript denote the
times of the red sideband excitation ($R$), blue sideband excitation ($B$) and
carrier process ($C$). The subscript denotes the sideband excitations corresponding to the
frequency of the $i$th membrane.} \label{fig4}
\end{figure}


By using the similar steps as for generating the NOON state with
$N=2$, an arbitrary NOON state of two mechanical resonators can also be generated by using the
steps schematically shown in Fig.~\ref{fig4}. The detailed steps are
summarized as below. In the step (i), by alternatively applying a series of the red-sideband excitations and carrier processes, we can prepare the state
\begin{equation}
\left\vert \psi \left( t_{1}\right)\right\rangle =\left\vert e,\frac{N}{2}-1,\frac{N}{2}%
\right\rangle
\end{equation}%
for the even number $N$ or
\begin{equation}
\left\vert \psi \left( t_{1}\right)\right\rangle =\left\vert e,\frac{N-1}{2},\frac{N-1}{2}%
\right\rangle
\end{equation}%
for the odd number $N$.

In the step (ii), for the even number $N$, the system is
driven with the red-sideband process $R_{1}$ for the membrane one with the time duration $\Delta t_{2}=\pi /4\Omega _{N/2}^{1}$, then we have
\begin{equation}
\left\vert \psi \left( t_{2}\right)\right\rangle =\frac{1}{\sqrt{2}}\left( \left\vert g,\frac{N%
}{2},\frac{N}{2}\right\rangle +\left\vert e,\frac{N}{2}-1,\frac{N}{2}%
\right\rangle \right).
\end{equation}
For the odd number $N$, the system is driven with the red sideband
excitation $R_{2}$ for the membrane two with the time duration $\Delta t_{2}=\pi /4\Omega _{(N+1)/2}^{2}$,
then we have
\begin{equation}
\left\vert \psi \left( t_{2}\right)\right\rangle =\frac{1}{\sqrt{2}}\left( \left\vert e,\frac{%
N-1}{2},\frac{N-1}{2}\right\rangle +\left\vert g,\frac{N-1}{2},\frac{N+1}{2}%
\right\rangle \right).
\end{equation}%

In the step (iii), for the even number $N$, we alternatively apply the driving field for $N/2$ and
$(N/2)-1$ red sideband excitations for membrane two and one with the
appropriate time durations, respectively, that is, an operation
operator $R_{2}(R_{1}R_{2})^{N/2-1}$ is acted on the state $
\left\vert \psi (t_{2})\right\rangle$, then we have
\begin{equation}\label{eq:66}
\left\vert \psi \left( t_{3}\right)\right\rangle =\frac{1}{\sqrt{2}}\left( \left\vert
e,N-1,0\right\rangle +\left\vert g,0,N\right\rangle \right).
\end{equation}
For the odd number $N$, $(N-1)/2$ red sideband excitations $R_{2}$ and
$R_{1}$ are applied to the membrane two and one, respectively, then
we also obtain the state shown in Eq.~(\ref{eq:66}).

\begin{figure}
\includegraphics[bb=10 15 370 270, width=8 cm, clip]{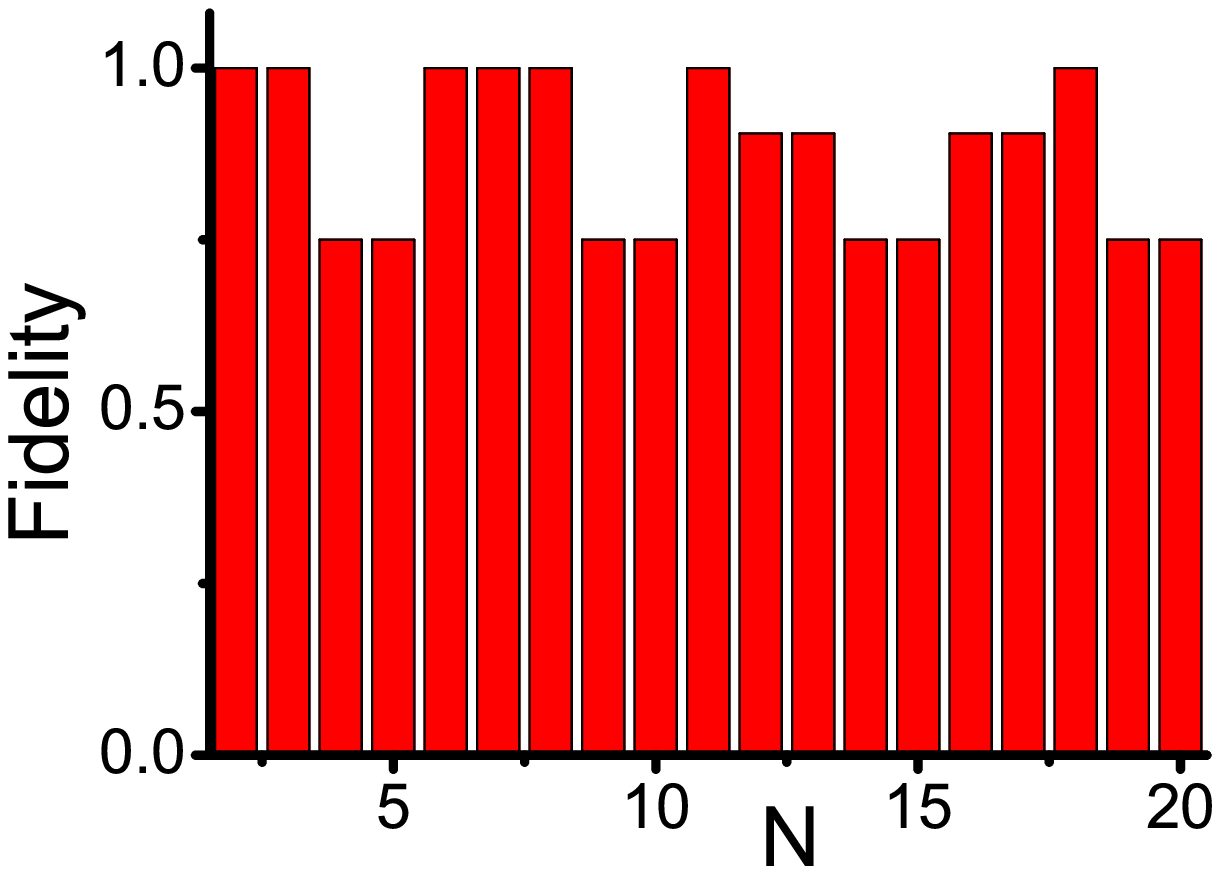}
\caption{Fidelity for preparing NOON state $\left(
\left\vert N,0\right\rangle +\left\vert 0,N\right\rangle \right)/\sqrt{2}$ from $N=2$ to $N=20$. } \label{fig5}
\end{figure}

In the step (iv), the system is driven with the red sideband excitation  $R_{1}$
for the membrane one with the time duration $\Delta t_{4}=\pi /2\Omega _{N}^{1}$, we have
\begin{equation}
\left\vert \psi \left( t_{4}\right)\right\rangle =|g\rangle\otimes\frac{1}{\sqrt{2}}\left( \left\vert
N,0\right\rangle +\left\vert 0,N\right\rangle \right).
\end{equation}
Thus the NOON state of the two mechanical resonators is prepared with the
cavity field in its ground state $\left\vert g\right\rangle $. We
note that there is information leakage in the third step because
the times to the two states $ \left\vert e,N-1,0\right\rangle$ and
$\left\vert g,0,N\right\rangle$ are not synchronized. The fidelity
of prepared NOON states from $N=2$ to $N=20$ due to such information
leakage is given in Fig.~\ref{fig5}, which shows that some
states cannot be prepared with the hundred percent. The detailed
discussion about this type of information leakage is given in the
Appendix B.

\subsection{Generating GHZ and W states of three mechanical modes}

We now turn to study the generation of the Greenberger-Horne-Zeilinger (GHZ) and W
states~\cite{Dur2000} of three vibrational membranes inside the cavity with the initial state $\left\vert \psi \left(t_{0}\right) \right\rangle =\left\vert g,0,0,0\right\rangle $. The definitions of the GHZ
state $\left\vert \psi \right\rangle _{\mathrm{GHZ}}$ and W state
$\left\vert \psi \right\rangle _{\mathrm{W}}$ are
\begin{equation}
\left\vert \psi \right\rangle _{\mathrm{GHZ}}=\frac{1}{\sqrt{2}}\left(
\left\vert 0,0,0\right\rangle +\left\vert 1,1,1\right\rangle \right).
\end{equation}%
and
\begin{equation}
\left\vert \psi \right\rangle _{\mathrm{W}}=\frac{1}{\sqrt{3}}\left(
\left\vert 1,0,0\right\rangle +\left\vert 0,1,0\right\rangle +\left\vert
0,0,1\right\rangle \right).
\end{equation}%

Let us first show the steps for preparing the W state. In
the step (i), the cavity field is driven by a laser field with the
carrier frequency. With an evolution time $\Delta t_{1}=\pi /2\Omega
$ and choosing the phase $\phi _{c}=3\pi /2$, the state of the
system becomes
\begin{equation}
\left\vert \psi \left( t_{1}\right) \right\rangle =\left\vert
e,0,0,0\right\rangle.
\end{equation}
In the step (ii),  the frequency of the driving field is tuned to
the red sideband excitation corresponding to frequency of the membrane one such that $\omega
_{d}=\omega_0 -\omega _{1}$. With the time duration $\Delta
t_{2}=[\arcsin \left( 1/\sqrt{3}\right)] /\Omega _{1}^{1}$ and the
phase $\phi _{r}^{1}=\pi /2$, the state of the system evolves to
\begin{equation}
\left\vert \psi \left( t_{2}\right) \right\rangle =\frac{1}{\sqrt{3}}%
\left\vert g,1,0,0\right\rangle +\sqrt{\frac{2}{3}}\left\vert
e,0,0,0\right\rangle.
\end{equation}
In the step (iii), the frequency of the driving field is tuned to
the red sideband excitation corresponding to the membrane two such that $\omega
_{d}=\omega_0 -\omega _{2}$. With the time duration $\Delta
t_{3}=\pi /4\Omega _{1}^{2}$ and the phase $\phi_{r}^{2}=\pi /2$,
the state of the system becomes
\begin{equation}
\left\vert \psi \left( t_{3}\right) \right\rangle =\frac{1}{\sqrt{3}}\left(
\left\vert g,1,0,0\right\rangle +\left\vert g,0,1,0\right\rangle +\left\vert e,0,0,0\right\rangle \right).
\end{equation}%
In the step (iv), the frequency of the driving field is tuned to the
red sideband excitation corresponding to the membrane three such that $\omega
_{d}=\omega_0 -\omega _{3}$. With the time duration $\Delta
t_{4}=\pi /2\Omega _{1}^{3}$ and the phase $\phi _{r}^{3}=\pi /2$,
the state of the system evolves to
\begin{equation}
\left\vert \psi \left( t_{4}\right) \right\rangle =|g\rangle\otimes\frac{1}{\sqrt{3}} \left(
\left\vert 1,0,0\right\rangle +\left\vert 0,1,0\right\rangle + \left\vert 0,0,1\right\rangle \right).
\end{equation}%
Thus the system is prepared in a product state of the W state of three
mechanical modes and the ground state $\left\vert g\right\rangle $
of the cavity field.

\begin{figure}
\includegraphics[bb=20 570 527 683, width=9 cm, clip]{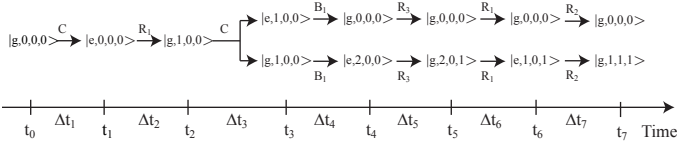}
\caption{Schematic diagram for preparing GHZ state $\left(
\left\vert 0,0,0\right\rangle +\left\vert 1,1,1\right\rangle
\right)/\sqrt{2}$ of three mechanical resonators. The
numbers in the superscript denote the
times of the red sideband excitation ($R$), blue sideband excitation ($B$) and
carrier process ($C$). The subscript denotes the sideband excitations corresponding to the
frequency of the $i$th membrane.} \label{fig6}
\end{figure}

Now, we are going to describe the steps of preparing the GHZ state,
as schematically shown in Fig.~\ref{fig6}. In the step (i), the cavity is driven with
the carrier frequency. With the evolution time $\Delta t_{1}=\pi
/2\Omega $ and the phase $\phi_{c}=3\pi /2$, the system
evolves to
\begin{equation}
\left\vert \psi \left( t_{1}\right) \right\rangle =\left\vert
e,0,0,0\right\rangle.
\end{equation}

In the step (ii), the frequency of the driving field is tuned to the
red sideband excitation corresponding to the membrane one such that $\omega
_{d}=\omega_0 -\omega _{1}$. With the time duration $\Delta
t_{2}=\pi /2\Omega _{1}^{1}$ and the phase $\phi _{r}^{1}=\pi /2$,
the state of the system evolves to
\begin{equation}
\left\vert \psi \left( t_{2}\right) \right\rangle =\left\vert
g,1,0,0\right\rangle.
\end{equation}

In the step (iii), the cavity field is driven again for the carrier
process such that $\omega _{d}=\omega_0 $. Then, with the time
duration $\Delta t_{3}=\pi /4\Omega $ and the phase $\phi _{c}=3\pi
/2$, the state of the system evolves to
\begin{equation}
\left\vert \psi \left( t_{3}\right) \right\rangle =\frac{1}{\sqrt{2}}%
\left\vert g,1,0,0\right\rangle +\frac{1}{\sqrt{2}}\left\vert
e,1,0,0\right\rangle.
\end{equation}

In the step (iv), the frequency of the driving field is tuned to the
blue sideband excitation corresponding to the frequency of the membrane one such that $\omega
_{d}=\omega_0 +\omega _{1}$. For the time duration $\Delta t_{4}$
satisfying the condition
\begin{equation}
    \sin(\Omega^{1}_{1} \Delta  t_{4})=\pm 1, \sin(\Omega^{1}_{2} \Delta  t_{4})=\pm 1,
\end{equation}
with the infinite approximation (see the Appendix B),  the state of the
system evolves to
\begin{equation}
\left\vert \psi \left( t_{4}\right) \right\rangle =\frac{1}{\sqrt{2}}%
\left\vert g,0,0,0\right\rangle +\frac{1}{\sqrt{2}}\left\vert
e,2,0,0\right\rangle.
\end{equation}

In the step (v), the driving field to tuned to the red sideband
excitation corresponding to the membrane three such that $\omega _{d}=\omega_0
-\omega _{3}$. With the time duration $\Delta t_{5}=\pi
/2\Omega_{1}^{3}$ and the phase $\phi _{b}^{3}=\pi /2$, the state of
the system evolves to
\begin{equation}
\left\vert \psi \left( t_{5}\right) \right\rangle =\frac{1}{\sqrt{2}}%
\left\vert g,0,0,0\right\rangle +\frac{1}{\sqrt{2}}\left\vert
g,2,0,1\right\rangle.
\end{equation}%

In the step (vi), the driving field to tuned the red sideband
excitation corresponding to the membrane one such that $\omega _{d}=\omega_0 -\omega
_{1}$.  With the time duration $\Delta t_{6}=\pi /2\Omega_{2}^{1}$
and the phase $\phi _{b}^{1}=3\pi /2$, the state of the system
evolves to
\begin{equation}
\left\vert \psi \left( t_{6}\right) \right\rangle =\frac{1}{\sqrt{2}}%
\left\vert g,0,0,0\right\rangle +\frac{1}{\sqrt{2}}\left\vert
e,1,0,1\right\rangle.
\end{equation}%

In the step (vii),  the driving field is tuned to the red sideband
excitation corresponding to the membrane two such that $\omega _{d}=\omega_0 -\omega
_{2}$. With the time duration $\Delta t_{7}=\pi /2\Omega_{1}^{2}$
and the phase $\phi _{b}^{1}=\pi /2$, the state of the system evolves to
\begin{equation}
\left\vert \psi \left( t_{7}\right) \right\rangle =|g\rangle\otimes\frac{1}{\sqrt{2}}
\left( \left\vert 0,0,0\right\rangle +\left\vert
1,1,1\right\rangle \right).
\end{equation}
Thus the whole system is prepared to a product state of the GHZ
state of three mechanical resonators and the ground state $\left\vert g\right\rangle $ of the
cavity field.

In principle, our method can be generalized to generation of the W
and GHZ states of $N$ membranes by sequentially applying a
series of red-sideband excitations, blue-sideband excitations and
carrier process with well chosen the time intervals, phases of the driving field. We have shown
the steps for the generation of those states in the appendix A.

\section{Preparation of entangled states beyond the Lamb-Dicke approximation}

In the above, we have studied a detailed method of generating entangled states in
the Lamb-Dicke approximation with $\eta _{i}\ll 1$. However, in some
optomechanical systems, e.g., a Bose-Einstein condensate served as the mechanical oscillator
coupled to the cavity field~\cite{Murch,Brennecke}, the condition $\eta _{i}\ll 1$ is
not satisfied, and also with the experimental progress, the
parameter $\eta _{i}$ outside the Lamb-Dicke regime is possible to be
achieved in the near future in the other types of optomechanical
systems. For the parameter $\eta _{i}$ outside the Lamb-Dicke approximation,
the higher-order powers of the Lamb-Dicke parameter should be taken
into account. We now show how to generate entangled states beyond
the Lamb-Dicke approximation by using a similar but not same method given in
Ref.~\cite{SBZheng}. Because the generation of the Bell state and the
W state can use the same method as in the Lamb-Dicke approximation
described in Sec.~III, thus below we focus on the generation of the
NOON state and the GHZ state. 

Beyond the Lamb-Dicke regime by assuming $k_{i}=j_{i}+n_{i}$,  the
Hamiltonian in Eq.~(\ref{eq:29}) can be rewritten as
\begin{equation}
H_{\rm int}=\hbar \Omega \sigma _{+}e^{-i\left( \omega _{d}t+\phi
_{d}\right) }\prod_{i=1}^{N} e^{-\frac{1}{2}\eta _{i}^{2}} H_{i} +{\rm H.c.},
\end{equation}
with
\begin{equation}\label{eq:85}
H_{i}= \sum_{n_{i}=-\infty }^{+\infty } \sum_{j_{i}=\max
[0,-n_{i}]}^{+\infty }\frac{\alpha^{i}_{j_{i},n_{i}}\left( b_{i}^{\dag }\right)
^{j_{i}}\left( b_{i}\right) ^{j_{i}+n_{i}}}{j_{i}!\left(
j_{i}+n_{i}\right) !}
\end{equation}
where $\alpha^{i}_{j_{i},n_{i}}=\left( -1\right) ^{j_{i}+n_{i}} \eta
_{i}^{2j_{i}+n_{i}}$. The Hamiltonian
in the interaction picture is given by $V(t)=e^{iH_{0}t/\hbar
}H_{\rm int}e^{-iH_{0}t/\hbar }$, thus we have
\begin{equation}
V(t)=\hbar \Omega \sigma _{+}e^{-i\left(\Delta _{d}t+\phi
_{d}\right) }\prod_{i=1}^{N}e^{-\frac{1}{2}\eta _{i}^{2}}H_{i}(t)
+{\rm H.c.},
\end{equation}
with
\begin{equation}
H_{i}(t)=\sum_{n_{i}=-\infty }^{+\infty }\sum_{j_{i}=\max
[0,-n_{i}]}^{+\infty } \frac{\alpha^{i}_{j_{i},n_{i}}\left( b_{i}^{\dag }\right)
^{j_{i}}\left( b_{i}\right) ^{j_{i}+n_{i}}}{j_{i}!\left(
j_{i}+n_{i}\right) !} e^{-i n_{i}\omega _{i}t}
\end{equation}
here $\Delta _{d}=\omega _{d}-\omega_{0}$. We assume that the off-resonant transitions can be
neglected in the resonant or near-resonant driving condition. When the
driving field satisfies the condition $\omega _{d}=\omega_{0}-\sum_{i=1}^{N}n_{i}\omega _{i}$, we have
\begin{equation}\label{eq:88}
V^{\left\{ n_{i}\right\} }=\hbar \Omega \sigma _{+}e^{-i\phi
_{d}}\prod_{i=1}^{N}e^{-\frac{1}{2}\eta _{i}^{2}}H_{i,n_{i}} +{\rm H.c.},
\end{equation}%
with $H_{i,n_{i}}$ given by
\begin{equation}
H_{i,n_{i}}= \sum_{j_{i}=\max
[0,-n_{i}]}^{+\infty }\frac{\alpha^{i}_{j_{i},n_{i}}\left( b_{i}^{\dag }\right)
^{j_{i}}\left( b_{i}\right) ^{j_{i}+n_{i}}}{j_{i}!\left(
j_{i}+n_{i}\right) !},
\end{equation}

\begin{figure}[t]
\includegraphics[bb=0 185 595 580, width=8.5 cm, clip]{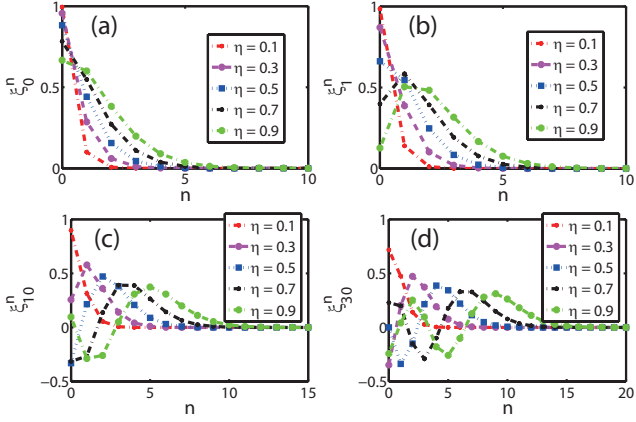}
\caption{$\xi_{m_{i}}^{n_{i}}$ in Eq.~(\ref{eq:95}) is plotted as the
function of $n_{i}$ for different values of $\eta =0.1,\, 0.3,\, 0.5,\, 0.7,\,
0.9$ in (a) $m=0$; (b) $m=1$; (c) $m=10$; and (d) $m=30$. Here, for the convenience,
we assume $n_{i}\equiv n$ and $m_{i}\equiv m$}
\label{fig7}
\end{figure}

With the Hamiltonian in Eq.~(\ref{eq:88}) and using the completeness relation in
Eq.~(\ref{eq:34}), the time evolution operator can be given as
\begin{equation}
U^{\left\{ n_{i}\right\} }\left( t\right) =\sum_{\left\{
m_{i}\right\} =0}^{+\infty }U_{\left\{ m_{i}\right\} }^{\left\{
n_{i}\right\} }\left( t\right),
\end{equation}%
where
\begin{equation}
U_{\left\{ m_{i}\right\} }^{\left\{ n_{i}\right\} }\left( t\right)
=\sum_{s=g}^{e}U^{\left\{ n_{i}\right\} }\left( t\right) \left\vert
s,\left\{ m_{i}\right\} \right\rangle \left\langle s,\left\{
m_{i}\right\} \right\vert.
\end{equation}%
Hereafter $\{ n_{i}\}$ denotes a number series, that is,  $\{ n_{i}\}\equiv
n_{1},n_{2},\cdots,n_{N}$. The operator $U_{\left\{ m_{i}\right\} }^{\left\{n_{i}\right\}
}\left( t\right) $ can be further written as
\begin{eqnarray}\label{eq:91}
U_{\left\{ m_{i}\right\} }^{\left\{ n_{i}\right\} }\left( t\right)
&=&\left( 1-\left\vert C_{\left\{ m_{i}-n_{i}\right\} }^{\left\{
n_{i}\right\} }\right\vert ^{2}\right) ^{1/2}\left\vert g,\left\{
m_{i}\right\} \right\rangle \left\langle g,\left\{ m_{i}\right\}
\right\vert
\notag \\
&&+C_{\left\{ m_{i}-n_{i}\right\} }^{\left\{ n_{i}\right\}
}\left\vert e,\left\{ m_{i}-n_{i}\right\} \right\rangle \left\langle
g,\left\{
m_{i}\right\} \right\vert   \notag \\
&&+\left( 1-\left\vert \widetilde{C}_{\left\{ m_{i}\right\}
}^{\left\{ n_{i}\right\} }\right\vert ^{2}\right) ^{1/2}\left\vert
e,\left\{ m_{i}\right\} \right\rangle \left\langle e,\left\{
m_{i}\right\} \right\vert
\notag \\
&&+\widetilde{C}_{\left\{ m_{i}\right\} }^{\left\{ n_{i}\right\}
}\left\vert g,\left\{ m_{i}+n_{i}\right\} \right\rangle \left\langle
e,\left\{ m_{i}\right\} \right\vert,
\end{eqnarray}%
where
\begin{equation}\label{eq:92}
C_{\left\{ m_{i}\right\} }^{\left\{ n_{i}\right\} }=-ie^{-i\phi
_{d}}\left( -1\right) ^{\sum_{i=1}^{N}n_{i}}\sin \left( \Omega
_{\left\{ m_{i}\right\} }^{\left\{ n_{i}\right\} }t\right)
\end{equation}%
with
\begin{equation}
\widetilde{C}_{\left\{ m_{i}\right\} }^{\left\{ n_{i}\right\}
}=-\left( C_{\left\{ m_{i}\right\} }^{\left\{ n_{i}\right\} }\right)
^{\ast },
\end{equation}%
and the Rabi frequency
\begin{equation}
\Omega _{\left\{ m_{i}\right\} }^{\left\{ n_{i}\right\} }=\Omega
\prod_{i=1}^{N}\xi _{m_{i}}^{n_{i}}.
\end{equation}%
If $n_{i}\geq 0$, $\xi _{m_{i}}^{n_{i}}$ is given as
\begin{equation}\label{eq:95}
\xi _{m_{i}}^{n_{i}}=e^{-\frac{1}{2}\eta _{i}^{2}}\eta _{i}^{n_{i}}\sqrt{%
\frac{m_{i}!}{\left( m_{i}+n_{i}\right) !}}L_{m_{i}}^{n_{i}}\left( \eta
_{i}^{2}\right)
\end{equation}%
and if $-m_{i}\leq n_{i}<0$, we have
\begin{eqnarray}\label{eq:96}
\xi _{m_{i}}^{n_{i}}&=&e^{-\frac{1}{2}\eta _{i}^{2}}(-\eta
_{i})^{-n_{i}}\sqrt{\frac{\left( m_{i}+n_{i}\right) !}{m_{i}!}}
L_{m_{i}+n_{i}}^{-n_{i}}\left( \eta _{i}^{2}\right) \nonumber\\
&=&(-1)^{|n_{i}|}\xi _{m_{i}-|n_{i}|}^{|n_{i}|},
\end{eqnarray}
here, $L_{m_{i}}^{n_{i}}\left( \eta_{i}^{2}\right)$ is the associated Laguerre
polynomials. Because $\xi _{m_{i}}^{n_{i}}$ is related to the effective
Rabi frequency, thus it is plotted as a function of $n_{i}$ in
Fig.~\ref{fig7}. We find that $\xi _{m_{i}}^{n_{i}}$ decreases
rapidly to the zero with the increase of $n_{i}$ when $\eta_{i} \ll 1$
thus the Lamb-Dicke approximation is valid in this regime. However,
with the increase of $\eta_{i}$, $\xi _{m_{i}}^{n_{i}}$ oscillates with
the increase of $n_{i}$ and finally approach to the zero. Thus if
$\eta_{i}$ is beyond the Lamb-Dicke regime, the terms for $n_{i}>1$ should
be taken into account and these terms will make the state
preparation much easier. It should be noted that $\xi
_{m_{i}}^{n_{i}}$ decreases to zero within a finite number $n_{i}$ even
$\eta_{i}=0.9$. That is, there is maximum phonon number that we can
create in one step even for the case that $\eta_{i}$ is beyond the
Lamb-Dicke regime.

We now study the preparation of the NOON state of two mechanical
resonators beyond the Lamb-Dicke regime. The time evolution operator
of the system for two membranes inside the cavity can be written out
from Eq.~(\ref{eq:91}) by taking the subscripts and the superscripts
$\{m_{i}\}\equiv m_{1}, m_{2}$, $\{m_{i}\pm n_{i}\}\equiv m_{1}\pm n_{1},
m_{2}\pm n_{2}$, and $\{n_{i}\}\equiv n_{1}, n_{2}$.
The coefficients in Eq.~(\ref{eq:92}) for the case that two
membranes inside the cavity are
\begin{equation}
C_{m_{1},m_{2}}^{n_{1},n_{2}}=-ie^{-i\phi _{d}}\left( -1\right)
^{n_{1}+n_{2}}\sin \left( \Omega
_{m_{1},m_{2}}^{n_{1},n_{2}}t\right)
\end{equation}%
with
\begin{equation}
\widetilde{C}_{m_{1},m_{2}}^{n_{1},n_{2}}=-\left(
C_{m_{1},m_{2}}^{n_{1},n_{2}}\right) ^{\ast }.
\end{equation}%
and the effective Rabi frequency
\begin{equation}
\Omega _{m_{1},m_{2}}^{n_{1},n_{2}} = \Omega \prod_{i=1}^{2}\xi _{m_{i}}^{n_{i}}.
\end{equation}
Here, $\xi _{m_{i}}^{n_{i}}$ is given by Eq.~(\ref{eq:95}) or
Eq.~(\ref{eq:96}). We assume that the system is initially in the state
$\left\vert \psi \left( t_{0}\right) \right\rangle =\left\vert
g,0,0\right\rangle $.

In the step (i), the cavity field is driven by
the external field which satisfies the carrier process. For the phase
$\phi _{d}=3\pi /2$ and the time duration $\Delta t_{1}=\pi /2\Omega _{0,0}^{0,0}$ with the evolution operator $U_{0,0}^{0,0}\left(
\Delta t_{1}\right)$, the state of the system becomes
\begin{equation}
\left\vert \psi \left( t_{1}\right) \right\rangle =\left\vert
e,0,0\right\rangle.
\end{equation}
In the step (ii), the frequency of the driving field is tuned to the
$N$th red sideband excitation corresponding to the frequency of the membrane one with $\Delta
_{d}=-N\omega _{1}$. With the time duration
$\Delta t_{2}=\pi /4\Omega _{0,0}^{N,0}$ and the phase $\phi
_{d}=\pi /2+N\pi$, the state of the system evolves to
\begin{equation}
\left\vert \psi \left( t_{2}\right) \right\rangle =\frac{1}{\sqrt{2}}%
\left\vert e,0,0\right\rangle +\frac{1}{\sqrt{2}}\left\vert
g,N,0\right\rangle.
\end{equation}
In the step (iii), the frequency of the driving field is tuned to the
$N$th red sideband excitation corresponding to the frequency of the membrane two such that $\Delta
_{d}=-N\omega _{2}$. With the time duration
$\Delta t_{3}=\pi /2\Omega _{0,0}^{0,N}$ and the phase $ \phi
_{d}=\pi /2+N\pi$, the state of the system evolves to
\begin{equation}
\left\vert \psi \left( t_{3}\right) \right\rangle =|g\rangle\otimes\frac{1}{\sqrt{2}}(
\left\vert N,0\right\rangle +\left\vert 0,N\right\rangle).
\end{equation}%
Thus the NOON state of two mechanical modes is prepared with the
cavity field in its ground state $\left\vert g\right\rangle $. It is
obvious that the preparation process is more efficient than the one
shown in the Lamb-Dicke regime.

We now show the steps of generating the GHZ state. The time evolution
operator of the three membranes inside the cavity can be
easily given from Eq.~(\ref{eq:91}) by taking the subscripts and the
superscripts as $\{m_{i}\}\equiv m_{1}, m_{2}, m_{3}$,
$\{m_{i}\pm n_{i}\}\equiv m_{1}\pm n_{1}, m_{2}\pm n_{2}, m_{3}\pm n_{3}$, and
$\{n_{i}\}\equiv n_{1}, n_{2}, n_{3}$.
The coefficients in Eq.~(\ref{eq:92}) for the case of the three
membranes are given as
\begin{equation}
C_{m_{1},m_{2},m_{3}}^{n_{1},n_{2},n_{3}}=\left[-ie^{-i\phi _{d}}\prod_{i=1}^{3}\left(
-1\right) ^{n_{i}}\right]\sin \left( \Omega
_{m_{1},m_{2},m_{3}}^{n_{1},n_{2},n_{3}}t\right)
\end{equation}%
and%
\begin{equation}
\widetilde{C}_{m_{1},m_{2},m_{3}}^{n_{1},n_{2},n_{3}}=-\left(
C_{m_{1},m_{2},m_{3}}^{n_{1},n_{2},n_{3}}\right) ^{\ast }
\end{equation}%
with the Rabi frequency
\begin{equation}
\Omega _{m_{1},m_{2},m_{3}}^{n_{1},n_{2},n_{3}} = \Omega \prod_{i=1}^{3}\xi _{m_{i}}^{n_{i}}.
\end{equation}
Here, $\xi _{m_{i}}^{n_{i}}$ is given by Eq.~(\ref{eq:95}) or
Eq.~(\ref{eq:96}).

We now assume that the system is initially prepared to the ground
state $ \left\vert \psi \left( t_{0}\right) \right\rangle
=\left\vert g,0,0,0\right\rangle $. In the step (i), the cavity
field is driven to the carrier process. With the interaction time
$\Delta t_{1}=\pi /4\Omega _{0,0,0}^{0,0,0}$ and the phase $\phi
_{d}=3\pi /2$, the state of the system evolves to
\begin{equation}
\left\vert \psi \left( t_{1}\right) \right\rangle
=\frac{1}{\sqrt{2}}\left( \left\vert g,0,0,0\right\rangle
+\left\vert e,0,0,0\right\rangle \right),
\end{equation}
with the time evolution operator $U_{0,0,0}^{0,0,0} \left( \Delta
t_{1}\right) $. In the step (ii),  the frequency of the driving field
is tuned to the red sideband excitation such that $\Delta_{d}=-\omega _{1}-\omega _{2}-\omega
_{3}$. With the time duration $\Delta t_{2}=\pi /2\Omega
_{0,0,0}^{1,1,1}$ and the phase $\phi_{d}=3\pi /2$, the state of the
system evolves to
\begin{equation}
\left\vert \psi \left( t_{2}\right) \right\rangle =|g\rangle\otimes\frac{1}{\sqrt{2}}%
(\left\vert 0,0,0\right\rangle +\left\vert 1,1,1\right\rangle).
\end{equation}
with the time evolution operator $U_{0,0,0}^{1,1,1}\left( \Delta
t_{2}\right) $. Thus the system is prepared to a product state of
the ground state $\left\vert g\right\rangle $ of  the cavity field and
the GHZ state of the three membranes.

\section{discussions and conclusions}

We discuss the experimental feasibility of the method by qualitatively considering
environmental effects and information leakage. (i)
The generation of entangled phonon states is based on the sideband
excitations, which are extensively used in the
optomechanical systems. Thus, our proposal should work in the resolved-sideband
regime which requires every frequency $\omega_{i}$ of the
vibrational mode of mechanical membrane is bigger than the decay rate $\gamma_{c}$ of the
cavity field, i.e., $\omega_{i}>\gamma_{c}$. (ii) The two-level approximation in our proposal
is guaranteed by the photon blockade effect, thus our method is more efficient when the
single-photon strong coupling strength $g_{i}$ is much bigger than
the decay rates $\gamma_{c}$ and $\gamma_{m,i}$ ($i=1, \,2, \cdots, N$)
of the cavity field and the mechanical modes, i.e., $g_{i} \gg
\gamma_{c},\,\gamma_{m,i}$.  Also Eq.~(\ref{eq:24}) shows that the more number of the mechanical resonators corresponds to the higher nonlinearity of the cavity field, thus corresponds to the better two-level approximation of the cavity field. (iii) During state preparation processes, negligible information leakage from the ground or the first excited state to other upper states of the cavity field requires that the anharmonicity  $ \sum_{i}(2 g_{i}^2/\omega_{i})$ of the
cavity field induced by the mechanical modes should be much bigger
than the strength  $\Omega$ of the classical driving field in the
carrier process~\cite{Xu_Liu}, i.e., $ \sum_{i}(2
g_{i}^2/\omega_{i})\gg\Omega$. (iv) To prevent information leakage due to
nearly resonant transitions induced by different mechanical resonators,
all of the transitions from the ground to the first excited state of the cavity field
induced by the driving field and different mechanical resonators should be well separated
in the frequency domain. In the
the Lamb-Dicke approximation, the frequency differences between any two of
membranes should satisfy the condition $|\omega_{i}-\omega_{j}|
\gg \Omega$. However, beyond the Lamb-Dicke approximation as shown in
Ref.~\cite{SBZheng}, the frequency differences between the membranes
should satisfy the conditions $|\omega_{i}-\omega_{j}|  \gg
\min(\omega_{i},\omega_{j}) \gg \Omega  \gg \gamma_{c},
\gamma_{m,i}$ or $\min(\omega_{i},\omega_{j}) \gg
|\omega_{i}-\omega_{j}| \gg \Omega   \gg \gamma_{c},
\gamma_{m,i}$. From (iv), we can find that the membrane number that
we can efficiently operate beyond the Lamb-Dicke
approximation is much smaller than that in the Lamb-Dicke approximation.

In summary, we have proposed a method to generate entangled states
of vibrational modes of multiple membranes inside the cavity via the
radiation pressure.  In particular, we carefully study the steps
for generating several typical entangled states, e.g., the Bell
and NOON state of two mechanical modes, the GHZ state and W
state of three mechanical modes for the parameters with and without the
Lamb-Dicke approximation.  We should emphasize as following.
(i) Basically, our method can be applied to other
optomechanical systems in which many mechanical modes are coupled to
a single-mode cavity field, such as optical cavity with levitating
dielectric microspheres~\cite{ChangPNAS,Romero-Isart10,Romero-Isart11,TCLi} or
trapped atomic ensembles~\cite{Brennecke,Murch}, optomechanical
crystals~\cite{Chan2009,Eichenfield09}, and microwave cavity with
nano-mechanical resonators~\cite{Massel}. (ii) Our proposal can be in
principle used to produce any kind of the entangled states. (iii) We
only qualitatively discuss the environmental effect and other information leakage on the
state preparation. The quantitative analysis on these factors will be
given in elsewhere. (iv) Our proposal is experimentally possible when the optomechanical
coupling strength approaches the single-photon strong coupling
limit.

\section{Acknowledgement}
YXL is supported by the National Natural Science Foundation of
China under Nos. 10975080 and 61025022.

\appendix
\begin{widetext}
\begin{figure*}
\includegraphics[bb=15 125 583 605, width=18 cm, clip]{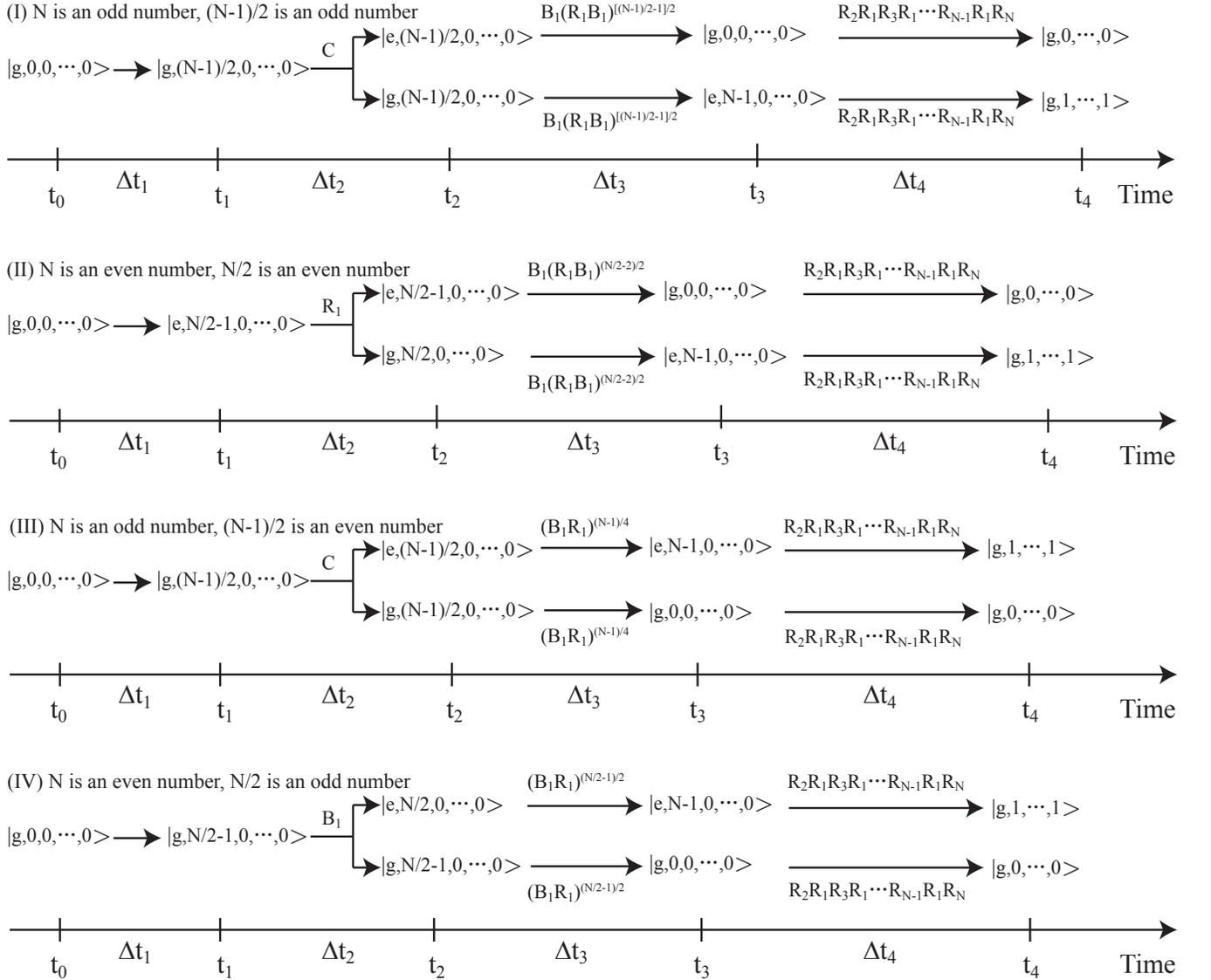}
\caption{Schematic diagrams for preparing GHZ states of $N$ membranes: (I) both $N$
and $(N-1)/2$ are odd numbers; (II) both $N$ and $N/2$ are even numbers;
(III) $N$ is an odd number and $(N-1)/2$ is an even number; (IV) $N$ is an even number and $N/2$
is an odd number.} \label{fig8}
\end{figure*}
\end{widetext}

\section{Preparation of W and GHZ states of $N$ membranes}

Follow the method we have given in Sec.~III, the W and GHZ states of $N$ membranes inside the cavity as
\begin{eqnarray}
\left\vert \psi \right\rangle _{\mathrm{W}}&=&\frac{1}{\sqrt{N}}
\left(\left\vert 1,0,0,\cdots ,0\right\rangle +\left\vert 0,1,0,\cdots,0\right\rangle \right. \nonumber \\
&& \left. \qquad+\cdots +\left\vert 0,0,\cdots ,0,1\right\rangle \right), \\
\left\vert \psi \right\rangle_{\mathrm{GHZ}}&=&\frac{1}{\sqrt{2}}\left( \left\vert 0\right\rangle
^{\otimes N}+\left\vert 1\right\rangle ^{\otimes N}\right),
\end{eqnarray}%
can be generated by sequentially applying a series of red-sideband excitations, blue-sideband excitations and carrier line operations. In this appendix, we are going to give the method for generating the W and GHZ states of $N$ membranes inside the cavity under the Lamb-Dicke approximation.

Assume that the system for $N$ membranes in the cavity is prepared in the state $\left\vert \psi \left( t_{0}\right) \right\rangle
=\left\vert g,0^{\otimes N}\right\rangle$ initially, then the W state of the $N$ membranes can be generated by one carrier process followed by $N$ red-sideband excitations. After the action of the carrier process for time duration $\Delta t_{1}=\pi /2\Omega$, the state of the system becomes
\begin{equation}
\left\vert \psi \left( t_{1}\right) \right\rangle =\left\vert e,0^{\otimes N}\right\rangle.
\end{equation}
at time $t_{1}=t_{0}+ \Delta t_{1}$. After that, $N$ red-sideband excitations (driving fields) are applied to the system sequentially, with frequencies $\omega _{d}=\omega_0 -\omega _{i}$, time durations $\Delta t_{i+1}=[\arcsin \left( 1/\sqrt{N+1-i}\right)] /\Omega _{1}^{i}$ and phase $\phi _{r}^{i}=\pi /2$ for $i=1,\cdots,N$. At time $t_{2}=t_{1}+\sum _{i=1}^{N} \Delta t_{i+1}$, the state of the system becomes
\begin{eqnarray}
\left\vert \psi \left( t_{2}\right)\right\rangle &=& |g\rangle \otimes \frac{1}{\sqrt{N}}
\left(\left\vert 1,0,0,\cdots ,0\right\rangle +\left\vert 0,1,0,\cdots
,0\right\rangle \right. \nonumber \\
&&\left. \qquad+\cdots +\left\vert 0,0,\cdots ,0,1\right\rangle \right).
\end{eqnarray}%

As shown in Fig.~\ref{fig8}, we can also generate the GHZ state of $N$ membranes for the system is initially prepared in the state
$\left\vert \psi \left( t_{0}\right) \right\rangle =\left\vert g,0^{\otimes
N}\right\rangle $. In step (i), by sequentially applying a series of red-sideband and carrier processes, we can prepare the state
\begin{equation}
\left\vert \psi \left( t_{1}\right) \right\rangle =\left\vert g,\frac{N-1}{2},0^{\otimes
\left( N-1\right) }\right\rangle
\end{equation}%
for the odd number $N$, or
\begin{equation}
\left\vert \psi \left( t_{1}\right) \right\rangle =\left\vert e,\left(
\frac{N}{2}-1\right) ,0^{\otimes \left( N-1\right) }\right\rangle
\end{equation}%
for the even numbers $N$ and $N/2$, or
\begin{equation}
\left\vert \psi \left( t_{1}\right) \right\rangle =\left\vert g,\left(
\frac{N}{2}-1\right) ,0^{\otimes \left( N-1\right) }\right\rangle
\end{equation}%
for the even number $N$ and odd number $N/2$.

In step (ii), if $N$ is an odd number, the system can be prepared in the state
\begin{equation}
\left\vert \psi \left( t_{2}\right) \right\rangle =\frac{1}{\sqrt{2}}\left( \left\vert
g\right\rangle +\left\vert e\right\rangle \right)\otimes \left\vert \frac{N-1}{2},0^{\otimes \left( N-1\right)
}\right\rangle,
\end{equation}
by the action of the carrier process for time duration $\Delta t_{2}=\pi /4\Omega$;
if both $N$ and $N/2$ are even numbers, the system can be prepared in the state
\begin{equation}
\left\vert \psi \left( t_{2}\right) \right\rangle =\frac{1}{\sqrt{2}}\left( \left\vert
e, \frac{N}{2}-1 \right\rangle + \left\vert g,\frac{N}{2}\right\rangle\right) \otimes \left\vert 0^{\otimes \left( N-1\right)
}\right\rangle,
\end{equation}
by the red sideband excitation of the membrane one for time duration $\Delta t_{2}=\pi /4\Omega^{1}_{N/2}$,
and if $N$ is an even number but $N/2$ is an odd number, the system can be prepared in the state
\begin{equation}
\left\vert \psi \left( t_{2}\right) \right\rangle =\frac{1}{\sqrt{2}}\left(\left\vert e,
\frac{N}{2}\right\rangle + \left\vert g,\frac{N}{2}-1\right\rangle\right) \otimes \left\vert
0^{\otimes \left( N-1\right) }\right\rangle,
\end{equation}
by the blue sideband excitation of the membrane one for time duration $\Delta t_{2}=\pi /4\Omega^{1}_{N/2}$.

In step (iii), we can prepare the state
\begin{equation}
\left\vert \psi \left( t_{3}\right) \right\rangle =\frac{1}{\sqrt{2}}\left( \left\vert
g,0\right\rangle +\left\vert e, N-1
\right\rangle \right)\otimes \left\vert 0^{\otimes \left( N-1\right) }\right\rangle
\end{equation}
for all the cases in step (ii) by a series of processes as shown in Fig.~\ref{fig8}. There are some information leaked in this step for the states $\left\vert g,0,0^{\otimes \left( N-1\right) }\right\rangle$ and $\left\vert e, N-1
,0^{\otimes \left( N-1\right) }\right\rangle$ can not be prepared in synchronism, and the fidelity is shown in Fig.~\ref{fig9} (discussion see the Appendix B).

In step (iv), by the action of $R_{2}R_{1}R_{3}R_{1}\cdots R_{N-1}R_{1}R_{N}$ for time duration $\Delta t_{4}=\sum _{i=2}^{N-1}\pi /2\Omega^{1}_{i}+\sum _{i=2}^{N}\pi /2\Omega^{i}_{1}$, we get the state
\begin{equation}
\left\vert \psi \left( t_{4}\right) \right\rangle =\left\vert g\right\rangle \otimes \frac{1}{%
\sqrt{2}}\left( \left\vert 0^{\otimes N}\right\rangle +\left\vert
1^{\otimes N}\right\rangle \right).
\end{equation}%
Thus we have prepared $N$ membranes in the GHZ state and the optical field in the cavity in the ground state $\left\vert g\right\rangle$.

\begin{figure}
\includegraphics[bb=10 15 370 270, width=8 cm, clip]{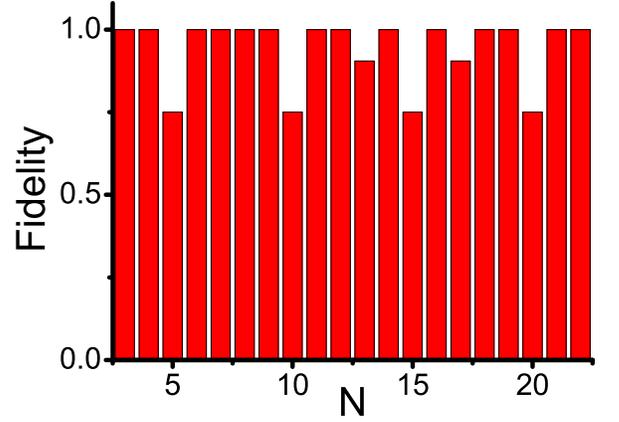}
\caption{Fidelity for preparing GHZ state $\frac{1}{
\sqrt{2}}\left( \left\vert 0^{\otimes N}\right\rangle +\left\vert
1^{\otimes N}\right\rangle \right)$ from $N=3$ to $N=22$.} \label{fig9}
\end{figure}

\section{The information leakage caused by non-synchronization}
In the preparation of the NOON state and GHZ state in the Lamb-Dicke regime, we need the processes for preparation two different states synchronically. For example, in the preparation of state $\left(\left\vert 2,0\right\rangle +\left\vert 0,2\right\rangle \right)/\sqrt{2}$, the transitions
$\left\vert g,1,1\right\rangle \rightarrow \left\vert e,1,0\right\rangle$, and
$\left\vert e,0,1\right\rangle \rightarrow \left\vert g,0,2\right\rangle$ should be synchronized.
In the other words, the time duration should satisfy Eq.~(\ref{eq:62}).
However, we find that Eq.~(\ref{eq:62}) can only be satisfied in the infinite approximation. In this appendix, we will discuss the information leakage caused by this type of non-synchronization.

Without lose of generality, suppose that the state of the system at time $t$ is given as
\begin{equation}\label{eq:A1}
\left\vert \psi \left( t\right) \right\rangle =\frac{1}{\sqrt{2}}%
\left\vert e,n-1,n'\right\rangle +\frac{1}{\sqrt{2}}\left\vert
g,m,m'\right\rangle
\end{equation}
and at the following time $t'$, we need to prepare the state
\begin{equation}\label{eq:A2}
\left\vert \psi \left( t'\right) \right\rangle =\frac{1}{\sqrt{2}}%
\left\vert g,n,n'\right\rangle +\frac{1}{\sqrt{2}}\left\vert
e,m-1,m'\right\rangle.
\end{equation}
So we drive the optical cavity with a laser field resonant to the first red sideband of the
first membrane for time duration $\Delta t= t'-t$, and the system evolves into
\begin{eqnarray}\label{eq:A3}
\left\vert \psi \left( t'\right) \right\rangle
&=&\frac{1}{\sqrt{2}}[ \cos \left( \Omega _{n}^{1} \Delta t\right) \left\vert e,n-1,n'\right\rangle  \nonumber \\
&&-ie^{i\phi _{r}^{1}}\sin \left( \Omega _{n}^{1} \Delta t\right) \left\vert g,n,n'\right\rangle \nonumber \\
&&+\cos \left( \Omega_{m}^{1} \Delta t\right) \left\vert g,m,m'\right\rangle  \nonumber \\
&&-ie^{-i\phi _{r}^{1}}\sin \left( \Omega _{m}^{1} \Delta t\right)\left\vert e,m-1, m'\right\rangle ].
\end{eqnarray}
where $\Omega _{n}^{1}=\Omega \eta _{1}\sqrt{n}$, $\Omega _{m}^{1}=\Omega \eta _{1}\sqrt{m}$.

In order to make sure that the system at the time $t'$ is in the state given by Eq.(\ref{eq:A2}), the time duration $\Delta t$ should satisfy the equations
\begin{equation}\label{eq:A4}
    \sin \left( \Omega _{n}^{1} \Delta t\right) = \pm 1, \sin \left( \Omega _{m}^{1} \Delta t\right)= \pm 1.
\end{equation}
Eq.(\ref{eq:A4}) can be equal to the equations
\begin{equation}\label{eq:A5}
    (\Omega _{n}^{1} \pm \Omega _{m}^{1})\Delta t= p \pi,
    \Omega _{m}^{1} \Delta t = (q + \frac{1}{2}) \pi,
\end{equation}
where $p$ and $q$ are positive integral numbers. Taking $\Omega _{n}^{1}/\Omega _{m}^{1}=\sqrt{n/m}$ (suppose $n \geq m$),
Eq.(\ref{eq:A5}) can be rewritten as
\begin{equation}\label{eq:A6}
    \left(\sqrt{\frac{n}{m}} \pm 1 \right)\left(q + \frac{1}{2}\right)= p.
\end{equation}
for choosing appreciate integral numbers $p$ and $q$.

\begin{figure}
\includegraphics[bb=15 15 380 295, width=8 cm, clip]{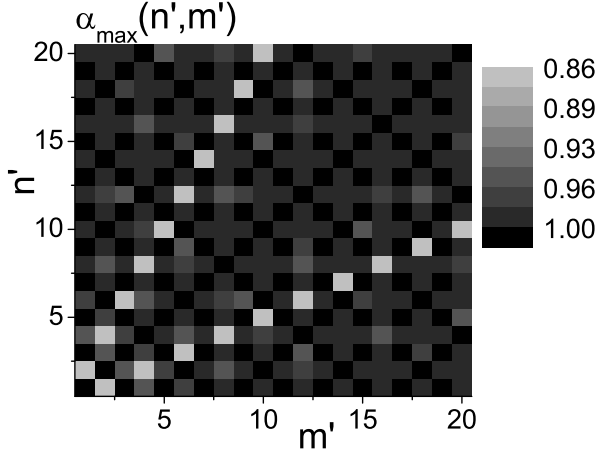}
\caption{The maximal value of $\alpha$ versus $n'$ and $m'$.} \label{fig10}
\end{figure}

According to the rate of the two angular frequencies, the solutions can be divided into three cases:

(i) If $\sqrt{n/m}$ is a rational number, we can rewrite it as a quotient of integers,
\begin{equation}\label{eq:A7}
    \sqrt{\frac{n}{m}}=\frac{n'}{m'},
\end{equation}
where $n'$ and $m'$ are integers and they have no common factors. Substitute Eq.(\ref{eq:A7}) into Eq.(\ref{eq:A6}), we get
\begin{equation}\label{eq:A8}
    \left(\frac{n' \pm m'}{m'} \right)\left( \frac{2q +1}{2}\right)= p.
\end{equation}
If both $n'$ and $m'$ are odd, then there are integral numbers $p$ and $q$ that satisfy Eq.(\ref{eq:A8}) exactly.

(ii) If $\sqrt{n/m}=n'/m'$ is a rational number, and one of them ($n'$ and $m'$) is even, there are no integral numbers satisfying Eq.(\ref{eq:A8}). However, there are integral numbers can satisfy the following equations for appreciate value of $\Delta t$,
\begin{equation}\label{eq:A9}
    \sin \left( \Omega _{n}^{1} \Delta t\right) = \pm \sin \left( \Omega _{m}^{1} \Delta t\right)= \pm  \alpha,
\end{equation}
where $\alpha$ is given by
\begin{equation}\label{eq:A10}
    \alpha=\left|\sin \left( \frac{m'}{n' \pm m'} p \pi \right)\right|.
\end{equation}
By the increase of $p$, $\alpha$ oscillates periodically with a period $n' \pm m'$. The maximal value of $\alpha$ as function of $n'$ and $m'$ (denoted by $\alpha_{max}(n',m')$) is shown in Fig.\ref{fig10}. From the figure we can see that $\alpha_{max}(n',m') \geq 0.86$, and $\alpha_{max}(n',m') = 1$ for the case that both $n'$ and $m'$ are odd, which is agree with the result given above.



(iii) If $\sqrt{n/m}$ is an irrational number, there are no integral numbers ($p$ and $q$) that satisfy Eq.(\ref{eq:A6}) exactly, but there is a rational number $n'/m'$ (both $n'$ and $m'$ are odd) that can be infinitely close to $\sqrt{n/m}$.
So we can get appreciate value of $\Delta t$ which can satisfy Eq.(\ref{eq:A6}) in the infinite approximation.

\bibliographystyle{apsrev}
\bibliography{ref}

\end{document}